\documentclass[journal=jacsat,manuscript=article]{achemso}
\setkeys{acs}{maxauthors=0}
\SectionNumbersOn 

\usepackage{amsfonts}
\usepackage{amssymb}
\usepackage{physics}
\usepackage{braket}
\usepackage{qcircuit}
\usepackage{graphicx}
\usepackage{dcolumn}
\usepackage{bm}
\usepackage[hidelinks]{hyperref}
\usepackage{tensor}
\usepackage{tabularx} 
\usepackage{booktabs}
\usepackage{multirow}
\usepackage{xurl}
\usepackage{xcolor}
\usepackage{chemformula}
\usepackage{threeparttable}
\usepackage[ruled, linesnumbered]{algorithm2e}

\newcommand{\rev}[1]{\textcolor{black}{#1}}
\newcommand{\revrev}[1]{\textcolor{black}{#1}}
\usepackage{qcircuit}

\title{A Further Comparison of \rev{MPS} and \rev{TTNS} for Nonadiabatic Dynamics of Exciton Dissociation}
\newcommand{\Aggre}{Guangdong Basic Research Center of Excellence for Aggregate Science, School of Science and Engineering, The Chinese University of Hong Kong, Shenzhen, Shenzhen, Guangdong, 518172, P.R. China}

\author{Weitang Li}
\email{liwt31@gmail.com}
\affiliation{\Aggre}

\author{Jiajun Ren}
\affiliation{MOE Key Laboratory of Theoretical and Computational
Photochemistry, College of Chemistry, Beijing Normal University,
Beijing, 100875, P. R. China}

\author{Jun Yan}
\affiliation{\Aggre}

\begin{document}

\begin{abstract}
Tensor networks, such as \rev{matrix product states (MPS)} and \rev{tree tensor network states (TTNS)}, are powerful \rev{ansätze} for simulating quantum dynamics. While both \rev{ansätze} are theoretically exact in the limit of large bond dimensions, [\textit{J. Chem. Theory Comput.} \textbf{2024}, \textit{20}, 8767–8781] reported a \rev{non-negligible} discrepancy in its calculations for exciton dissociation. To resolve this inconsistency, we conduct a systematic comparison using Renormalizer, a unified software framework for MPS and TTNS.
By revisiting the benchmark P3HT:PCBM heterojunction model, we show that the observed discrepancies arise primarily from insufficient bond dimensions. 
By increasing bond dimensions, \revrev{we reduce the relative difference in occupancy for
   weakly populated electronic states from up to 60\% towards the end of the
   simulation to less than 10\% and the absolute difference from 0.05 to
   0.005.}
\rev{We also discuss the impact of tensor network structures on accuracy and efficiency, with the difference further reduced by an optimized TTNS structure.}
Our results confirm that both methods converge to numerically exact solutions when bond dimensions are adequately scaled. This work not only validates the reliability of both methods but also provides high-accuracy benchmark data for future developments in quantum dynamics simulations.

\end{abstract}

\maketitle

\section{Introduction}
Tensor networks have emerged as powerful tools for simulating quantum dynamics. 
These networks approximate high-order wavefunction tensors as contractions of lower-order tensors, enabling accurate simulations of complex systems~\cite{orus2019tensor, banuls2023tensor, larsson2024tensor}. 
The accuracy of this approximation is controlled by the size of the low-order tensors, 
also known as the bond dimension. 
In the limit of large bond dimensions, tensor network methods converge to the exact solution.
Among tensor network methods, matrix product states (MPS) are the foundation of the time-dependent density matrix renormalization group (TD-DMRG)~\cite{schollwock2011density, paeckel2019time, ren2022time, haegeman2011time, lubich2015tt, lubich2015mctdh, haegeman2016unifying}. 
TD-DMRG has been successfully applied to study systems such as charge transport in organic semiconductors~\cite{li2021general}, non-radiative decay in molecular aggregates~\cite{wang2023minimizing}, and photoinduced ultrafast vibration-coupled electron transfer reactions~\cite{wang2025numerically}, among others~\cite{borrelli2017simulation, greene2017tensor, ren2018time, ma19, baiardi2019large, borrelli2021finite, peng2023studies}. 
Meanwhile, tree tensor network states (TTNS) form the basis of the multi-layer multiconfiguration time-dependent Hartree method (ML-MCTDH)~\cite{shi2006classical, wang2015multilayer}, which has found wide-spread applications~\cite{wang2008coherent, wang2010coherent, vendrell2011multilayer, weike2022multi, ke2023tree, reddy2024intramolecular, zheng2025ml} 
such as the study of singlet fission~\cite{schroder2019tensor} and exciton migration~\cite{binder2018conformational, binder2020first, di2020quantum, popp2021quantum, mondelo2022quantum, brey2024coherent}.

While TD-DMRG and ML-MCTDH originated in different research communities~\cite{meyer1990multi, manthe1992wave, white1992density, white1993density, wang2003multilayer,white2004real, daley2004time, vidal2004efficient, feiguin2005time}, 
it is now well-established that both rely on tensor networks and share fundamental similarities~\cite{larsson2024tensor}. 
Recent efforts have focused on directly comparing these methods to understand how their different tensor network structures affect simulation accuracy and efficiency~\cite{gunst2018t3ns, larsson2019computing, ren2022time, li2024optimal, larsson2024tensor, chen2025tree}.
In principle, both methods should converge to the exact result when the bond dimension is sufficiently large. 
It is often assumed that TTNS converges faster than MPS with respect to the bond dimension, while MPS offers a better computational scaling in terms of the bond dimension, due to its simpler tensor structure.

However, an \rev{intriguing} discrepancy emerged in a recent benchmark study~\cite{dorfner2024comparison}.
The authors performed an extensive and careful comparison of the two methods \revrev{with standard implementations}, and found that both methods agreed perfectly in most scenarios.
Their analysis also connected the observed dynamics with patterns of entanglement entropy and the underlying tensor network structures.
For exciton dissociation at poly(3-hexylthiophene):[6,6]-phenyl-\ch{C61}-butyric acid methyl ester (P3HT:PCBM) heterojunctions, TD-DMRG and ML-MCTDH showed \rev{small but non-negligible} differences \rev{at the long time limit}. 
This inconsistency is troubling because both methods are frequently used as reference standards for validating and benchmarking other quantum dynamics approaches~\cite{peng2022exciton, xu2022taming, li2023efficient, liu2024benchmarking}.
The comparison is complicated by the use of different software implementations~\cite{mctdh:MLpackage, fishman2022itensor},
\rev{making it challenging to isolate the origin of these differences}.
Very recently, Lindoy and Rungger \textit{et al.} systematically investigated this system using both multiset and singleset MPS/TTNS ansätze based on the pyTTN package~\cite{lindoy2025pyttn}. They demonstrated that the multiset ansätze yields highly accurate and consistent results, while the singleset ansätze exhibits slow bond dimension convergence. Their analysis suggests that insufficient bond dimension is the most likely source of the discrepancy reported in the benchmark study~\cite{dorfner2024comparison}.

In this work, we revisit the P3HT-PCBM model using the Renormalizer package,
which provides a unified framework for both MPS and TTNS~\cite{li2024optimal, renormalizer}.
\rev{Through extensive benchmark using the projector splitting (PS) algorithm based on the Time Dependent Variational Principle (TDVP)~\cite{haegeman2011time, lubich2015tt, lubich2015mctdh, haegeman2016unifying,lindoy2021time, lindoy2021time2, ceruti2021time}},
we conclusively confirm that the observed discrepancies stem from insufficient bond dimensions.
\revrev{In this context, TDVP-PS with MPS corresponds to TD-DMRG, while TDVP-PS with TTNS corresponds to a variant of ML-MCTDH known as Projector Splitting Integrator ML-MCTDH (PSI-ML-MCTDH)}.
\rev{
Additionally, guided by the entanglement entropy calculated from TTNS simulations, we propose two new structures for MPS and TTNS, respectively, tailored for this exciton dissociation model. We find that the new TTNS structure is more efficient than the previous MPS and TTNS structures, while the new MPS structure results in higher error.
}
Our results demonstrate that both methods can achieve numerically exact results, reinforcing their reliability for the simulation of complex quantum dynamics.

\section{Methodology}
\subsection{The Exciton Dissociation Model}
\label{sec:model}
\begin{figure}[ht]
    \centering    \includegraphics[width=.7\textwidth]{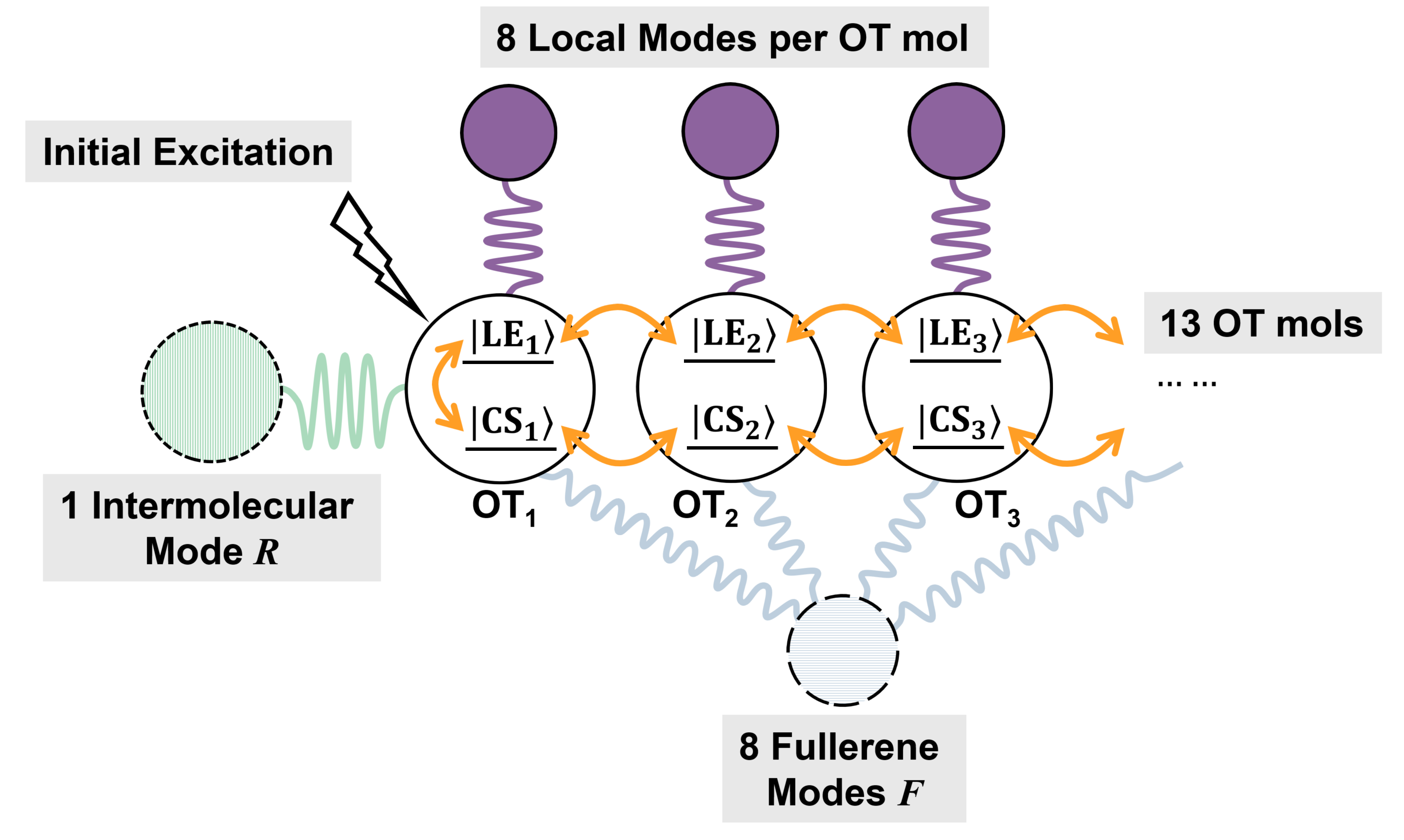}
    \caption{The exciton dissociation model employed in this work. Each OT molecule is associated with a LE and a CS state. The system consists of 13 OT molecules in total. Orange arrows indicate electronic state transitions, while the purple, green, and light blue spheres represent different types of vibrational modes coupled to the LE and CS states.}
    \label{fig:model}
\end{figure}

In this section,
we present the model for the numerical comparison of \rev{MPS and TTNS}.
The model describes the exciton dissociation at the interface of fullerene molecules and a linear chain of $N_o=13$ oligothiophene (OT) molecules, which is originally introduced in the reference paper~\cite{tamura2013ultrafast, huix2015concurrent, dorfner2024comparison}. 
The model can be considered as a simplified representation of the P3HT:PCBM heterojunction~\cite{Dang2011P3HT}.
The reference paper provides a detailed chemical picture of the model.
In Fig.~\ref{fig:model}, we present an interaction-based view of the model.

As shown in Fig.~\ref{fig:model}, each OT molecule is associated with two electronic states: a local excitation (LE) state and a charge-separated (CS) state. \rev{The CS state} describes the interaction between the OT molecule and the fullerene cluster. 
The LE and CS states can hop to their nearest neighbors. 
Additionally, the LE and CS states at the first OT molecule (LE$_1$ and CS$_1$) can transit between each other. 
The system starts in the LE$_1$ state due to initial excitation.
Besides, there are three types of vibrations in the model, which are shown as filled spheres in Fig.~\ref{fig:model}.
Each OT molecule has 8 local vibrational modes, and they are coupled to both its LE state and the CS state.
The fullerene cluster has 8 vibrational modes, denoted as $F$, which couple to all CS states.
There is an additional intermolecular vibrational mode, denoted as $R$, which specifically couples to the transition between LE$_1$ and CS$_1$.

The Hamiltonian of the system $\hat H$ can be decomposed into three parts
\begin{equation}
    \hat H = \hat H^{\textrm{e}} + \hat H^{\textrm{vib}} + \hat H^{\textrm{e-vib}} \ .
\end{equation}
Here, $\hat H^{\textrm{e}}$ represents the electronic energy and interactions, 
$\hat H^{\textrm{vib}} $ represents the vibration energy,
and $ \hat H^{\textrm{e-vib}} $ represents the electron-phonon interaction or the vibronic coupling.
We next describe each term in detail.

The electronic Hamiltonian $\hat H^{\textrm{e}}$ is further split into diagonal and off-diagonal terms
\begin{equation}
    \hat H^{\textrm{e}} = \hat H^{\textrm{e}}_{\textrm{diag}} + \hat H^{\textrm{e}}_{\textrm{off-diag}} \ .
\end{equation}
The diagonal part consists of on-site energies for the LE and CS states:
\begin{equation}
\label{eq:ham-e}
    \hat H^{\textrm{e}}_{\textrm{diag}} = \epsilon^{\textrm{LE}} \sum_n^{N_o} \ket{\textrm{LE}_n}\bra{\textrm{LE}_n}
     +  \sum_n^{N_o} \epsilon^{\textrm{CS}}_n \ket{\textrm{CS}_n}\bra{\textrm{CS}_n} \ . \\
\end{equation}
The off-diagonal terms include hopping interactions between neighboring LE and CS states:
\begin{equation}
     \hat H^{\textrm{e}}_{\textrm{off-diag}} = J \sum_n^{N_o-1} \ket{\textrm{LE}_n}\bra{\textrm{LE}_{+1}}
     +  t \sum_n^{N_o-1} \ket{\textrm{CS}_n}\bra{\textrm{CS}_{n+1}}  + \lambda \ket{\textrm{LE}_1}\bra{\textrm{CS}_{1}}
      + \textrm{H.C.}    
\end{equation}
\rev{where ``H.C.'' represents Hermitian conjugation.}
Following the reference work, the parameters used in this model are  $ \epsilon^{\textrm{LE}} = 100$ meV, 
$J = 100$ meV, $t = -120$ meV and $\lambda = -200$ meV.
The values for $\epsilon^{\textrm{CS}}_n$ are listed in Table~\ref{tab:epsilon_cs}.

The vibrational Hamiltonian $\hat H^{\textrm{vib}}$  includes three types of vibrations:
an intermolecular mode $R$, $N_f$ fullerene modes $F$ and $N_o\times N_m$ local OT vibrational modes.
When there is no ambiguity, we refer to the local OT vibrations simply as ``OT''.
$\hat H^{\textrm{vib}}$ is then written as
\begin{equation}
    \hat H^{\textrm{vib}} = \omega_R \hat b^\dagger_{R} \hat b_{R}
    + \sum_l^{N_f} \omega_{F, l} \hat b^\dagger_{F, l} \hat b_{F, l}
    + \sum_n^{N_o}\sum_l^{N_m} \omega_{\textrm{OT}, l} \hat b^\dagger_{\textrm{OT}, nl} \hat b_{\textrm{OT}, nl} \ .
\end{equation}
Here, $N_f=N_m=8$, $\omega_R = 10$ meV, and the frequencies $\omega_{F, l}$ and $\omega_{\textrm{OT}, l}$ are listed in Table~\ref{tab:vibrations} in \ref{sec:params}.

$\hat H^{\textrm{e-vib}}$ assumes linear vibronic coupling and
can be decomposed to three groups, corresponding to three different types of vibrations:
\begin{equation}
    \hat H^{\textrm{e-vib}} = \hat H^{\textrm{e-vib}}_{R} + \hat H^{\textrm{e-vib}}_{F} + \hat H^{\textrm{e-vib}}_{\textrm{OT}} \ .
\end{equation}
The intermolecular $R$ mode has both diagonal and off-diagonal couplings to the LE$_1$ state and CT$_1$ state
\begin{equation}
    \hat H^{\textrm{e-vib}}_{R} = g_R \left( \hat b^\dagger_{R} + \hat b_R \right) \ket{\textrm{CS}_1} \bra{\textrm{CS}_1} + 
    g'_R \left( \hat b^\dagger_{R} + \hat b_R \right) \left ( \ket{\textrm{LE}_1} \bra{\textrm{CS}_1} + \textrm{H.C.} \right ) \ ,
\end{equation}
where $g_R = -10/\sqrt{2}$ meV and $g'_R = -30/\sqrt{2}$ meV.
The fullerene modes $F$ couple to the on-site energy of the CS states
\begin{equation}
    \hat H^{\textrm{e-vib}}_{F} = \sum_n^{N_o}\sum_l^{N_f} g_{F, l} \left( \hat b^\dagger_{F, l} + \hat b_{F, l} \right) \ket{\textrm{CS}_n} \bra{\textrm{CS}_n} \ ,
\end{equation}
where the coupling constants $g_{F, l}$ are included in Table~\ref{tab:vibrations}.
The OT modes couple to the on-site energy of both the LE and CS states
\begin{equation}
    \hat H^{\textrm{e-vib}}_{\textrm{OT}} = \sum_n^{N_o}\sum_l^{N_f} \left [
    g_{\textrm{OT}, l} \left( \hat b^\dagger_{\textrm{OT}, nl} + \hat b_{\textrm{OT}, nl} \right) \ket{\textrm{CS}_n} \bra{\textrm{CS}_n}+ 
    g'_{\textrm{OT}, l}  \left( \hat b^\dagger_{\textrm{OT}, nl} + \hat b_{\textrm{OT}, nl} \right) \ket{\textrm{LE}_n} \bra{\textrm{LE}_n}
    \right] \ .
\end{equation}
The coupling constants $g_{\textrm{OT}, l}$ and $g'_{\textrm{OT}, l} $ are included in Table~\ref{tab:vibrations}.
These coupling constants, including $F$ and OT vibrations, are assumed to be independent of $n$, the index of the OT molecule.

Overall, the exciton dissociation model consists of 26 electronic states and $1+8+8\times13=113$ vibrational modes. 
While most of the interactions are short ranged, $\hat H^{\textrm{e-vib}}_{F}$ introduces long range interactions between fullerene and OT molecules.

\subsection{MPS and TTNS}
\label{sec:tn}
In this section, we briefly introduce the principles of MPS and TTNS, using the tensor network language and with an emphasis on the relationship between their accuracy and the entanglement entropy.
For more detailed explanations, readers are encouraged to consult several excellent reviews \rev{and introductory literatures}~\cite{beck2000multiconfiguration,schollwock2011density, orus2014practical, chan2016matrix, hauschild2018efficient, ren2022time, larsson2024tensor}.

MPS and TTNS are data structures that approximate high-order tensors by the contraction of low-order tensors.
Let us consider a quantum system with $N$ degrees of freedom.
For each degree of freedom, the corresponding primitive basis is denoted by $\ket{\sigma_i}$.
The wavefunction of the system can be approximated by MPS and TTNS as
\begin{equation}
\label{eq:ttns}
    \ket{\Psi} = \sum_{\{a\}, \{\sigma\}} A[1]^{\sigma_1}_{\Lambda_1, a_1} 
    A[2]^{\sigma_2}_{\Lambda_2, a_2} 
    \cdots
    A[N]^{\sigma_N}_{\Lambda_N, a_N} 
    \ket{\sigma_1 \sigma_2 \cdots \sigma_N} \ .
\end{equation}
Here $A[i]$ represents the low-order tensors.
$\Lambda_i$ is a set of indices that connects to child nodes and $a_i$ connects to the parent node. 
If $\Lambda_i$ is an empty set, then $A[i]$ represents a leaf node.
The contraction between the tensors occur according to the index $\Lambda_i$.
While TTNS employs a general tree-like contraction topology, MPS utilizes a linear chain structure, 
making MPS a special case of TTNS where all tensors are arranged in one dimension.
Eq.~\ref{eq:ttns} represents a general TTNS
and it reduces to a MPS if $\Lambda_i=\{a_{i-1}\}$.

Since tensor contraction operation is undirected, the concept of parent and children nodes in TTNS is arbitrary.
In other words, we are free to choose the root of the tree for implementation or formal purposes and it does not affect the contraction result in Eq.~\ref{eq:ttns}.
If a specific tensor is chosen as the root,
we denote it as $A[r]$, where the index $a_r$ is not included in any $\Lambda_i$
and the dimension of $a_r$ is 1.
As a result, the index $a_r$ is omitted in the following when no ambiguity arises.

In ML-MCTDH or three-legged tree tensor network~\cite{gunst2018t3ns}, there are entirely ``virtual'' nodes, which are not associated with any physical degree of freedom $\sigma$.
For these virtual nodes, an auxiliary physical degree of freedom with a Hilbert space of dimension 1 can be assigned 
to ensure consistency with Eq.~\ref{eq:ttns}.
Similarly, there are also nodes where the number of physical degree of freedom is greater than 1
and they can be combined so that formally only one physical degree of freedom presents, which is also known as the mode combination technique.
For simplicity, we assume that at the root node the dimension of $\sigma_r$ is 1 and thus $\sigma_r$ can be ignored.
If $\abs{\sigma_r}\neq 1$, we can decompose $A[r]^{\sigma_r}_{\Lambda_r}$ into two tensors.
One that has the indices $\Lambda_r$ acts as the new root, and the other whose shape is $\abs{\sigma_r}\times\abs{\sigma_r}$ acts as a child to the new root.
An auxiliary physical degree of freedom can then be assigned to the new root and $\abs{\sigma_r}$ becomes 1.
In this work $|\sigma|$ or $|a|$ represent the dimension of the index.

Through sequential QR or \rev{singular value decomposition}, the tensor network in Eq.~\ref{eq:ttns} can be transformed into a ``canonical'' format.
A tensor $A[i]$ is said to be canonical if it satisfies the following condition:
\begin{equation}
\label{eq:cano-condition}
    \sum_{\Lambda_i, \sigma_i} A^\dagger[i]^{\sigma_i}_{\Lambda_i, a'_i}A[i]^{\sigma_i}_{\Lambda_i, a_i} = \delta_{a'_ia_i} \ .
\end{equation}
Eq.~\ref{eq:ttns} is considered canonical if all tensors $A[i]$, except for the root $A[r]$, are canonical.
The root node is thus also referred to as the canonical center.
Importantly, the canonical center can be moved to any node in the tree, analogous to how any node can serve as the root.
In the canonical form, the wavefunction $\ket{\Psi}$ can be expressed as
\begin{equation}
\label{eq:cano-expansion}
    \ket{\Psi} = \sum_{\Lambda_r} A[r]_{\Lambda_r}\prod_{\{j;a_j \in \Lambda_r\}}\ket{a[j]_{a_j}} \ ,
\end{equation}
where $A[r]$ acts as the coefficient tensor, and $\{\ket{a[j]}\}$ forms an orthogonal basis set.
Here each $\ket{a[j]}$ is the orthogonal basis set by one of the children
\begin{equation}
    \ket{a[j]_{a_j}} = \sum_{\Lambda_j, \sigma_j} A[j]^{\sigma_j}_{\Lambda_j, a_j}\ket{\sigma_j}\prod_{\{k;a_k \in \Lambda_j\}}\ket{a[k]_{a_k}}
\end{equation}
The orthogonal relation $\braket{a[j]_{a'_j}|a[j]_{a_j}}=\delta_{a'_ja_j}$ can be derived from
the canonical condition Eq.~\ref{eq:cano-condition} through tree recursion.

The accuracy of the tensor network approximation is controlled by the dimension of the indices $\{a\}$.
In MPS and general tensor network literature, this quantity is typically referred to as the bond dimension. 
In the context of ML-MCTDH, the same quantity is called the number of single-particle functions.
In this work, we employ the term ``bond dimension'' following our previous conventions and use $M$ to denote this quantity.

In principle, the bond dimension can vary for each bond in the tree, and it can be adjusted dynamically during time evolution.
In this work, we employ the same fixed bond dimension for every bond, \rev{primarily to simplify  the setup.}
Since the initial state of the model introduced in Sec.~\ref{sec:model} is a product state,
Krylov subspace vectors $\hat H^n \ket{\Psi}$ are added to the product state with small coefficients to expand the bond dimension to the target value $M$.
\rev{
Specifically, we typically add around 10 such vectors, included together with a small weight of $10^{-10}$, to ensure minimal perturbation while effectively expanding the state space. This enrichment is performed only once during the initial state preparation and is not repeated during the time evolution. 
We note that here the Krylov vector refers to the global wavefunction, and should not be confused with the Krylov vector method employed for solving the matrix exponential of the time evolution discussed at the end of Sec.~\ref{sec:tns}.}

The error introduced by the finite bond dimension can be quantified by the singular values of $A[r]$.
\rev{The square of the singular values corresponds to the natural orbital population in MCTDH language.}
Suppose $a_j \in \Lambda_r$ and the other indices \rev{of the root node} are denoted as $\Lambda'_r=\Lambda_r \setminus \{a_j\}$,
we can reshape $A[r]_{\Lambda_r}$ to a two dimensional tensor, where the first and the second indices correspond to $\Lambda'_r$ and $a_j$ respectively.
By performing singular value decomposition on $A[r]_{\Lambda'_r, a_j}$,
we obtain the singular values $s_l$.
\rev{Let $M'$ be a integer smaller than $|a_j|$}.
By keeping only the first (largest) $M'$ singular values, the dimension of $a_j$ is compressed to $M'$
\begin{equation}
\label{eq:svd}
    A[r]_{\Lambda'_r, a_j} \stackrel{\text{SVD}}{=} \sum_l^{|a_j|} U_{\Lambda_rl} s_lV_{la_j} \approx \sum_l^{M'} U_{\Lambda_rl} s_lV_{la_j} \ .
\end{equation}
Let $\ket{\Psi'}$ be the wavefunction after compression. 
Since $U$ and $V$ are unitary matrices and the bases in Eq.~\ref{eq:cano-condition} are orthogonal,
the compression fidelity is
$
    \abs{\braket{\Psi'|\Psi}}^2 = \sum_l^{M'} s^2_l
$.
Assuming the wavefunction is normalized, $\sum_l^{|a_j|} s^2_l = 1$.
The compression error is the sum of the square of discarded singular values
\begin{equation}
\label{eq:compression-error}
    1 - \abs{\braket{\Psi'|\Psi}}^2 = \sum_{l={M'+1}}^{|a_j|} s^2_l \ .
\end{equation}
Thus, the ideal wavefunction for tensor network approximation are those whose singular values decay rapidly.

The efficiency of the tensor network compression is closely related to the bipartite von Neumann entanglement entropy $S$.
In a tree tensor network, cutting an arbitrary bond (edge) $a_k$ divides the system degrees of freedom into two parts $X$ and $Y$.
The bipartite von Neumann entanglement entropy $S$ is defined as
\begin{equation}
\label{eq:ee}
    S = -\Tr{\rho_X\ln{\rho_X}} = -\Tr{\rho_Y\ln{\rho_Y}}
\end{equation}
where $\rho_X$ and $\rho_Y$ are the reduced density matrix of $X$ and $Y$, respectively.

In general, $S$ is difficult to calculate for many-body systems,
because calculating $S$ involves \rev{diagonalizing} the reduced density matrices to obtain the eigenvalues.
However, the canonical form of tensor networks, as described in Eq.~\ref{eq:cano-expansion}, provides an efficient way to calculate $S$.
To do so, the canonical center is firstly moved to the bond $a_k$ that divides the sub-systems such that $a_k \in \Lambda_r$.
The remaining indices are denoted as $\Lambda'_r=\Lambda_r \setminus \{a_k\}$.
Supposing $X$ is in the subtree of the bond $a_k$,
its reduced density matrix $\rho_X$ is then
\begin{equation}
    \rho_X = \Tr_{Y}{\ket{\Psi}\bra{\Psi}} = \sum_{\Lambda'_r, a_k, a'_k} A^\dagger[r]_{\Lambda'_r, a'_k} A[r]_{\Lambda'_r, a_k} \ket{a[k]_{a_k}}\bra{a[k]_{a'_k}}
\end{equation}
By performing SVD on $A[r]$ as described earlier in Eq.~\ref{eq:svd},
the entanglement entropy can be calculated by the singular values
\begin{equation}
\label{eq:ee-s}
    S = -\sum_l s^2_l \ln s^2_l
\end{equation}
Note that SVD can also be employed to move the canonical center to the neighbouring nodes.
Thus, by sweep the canonical center across the tree, we can obtain the entanglement entropy $S$ for each bond.

As indicated by Eq.~\ref{eq:ee-s}, if the bond $a_k$ has bond dimension $M$, 
the maximumly possible entanglement entropy $S$ is $\ln M$, which occurs when $s_l = 1/\sqrt{M}$.
In other words, for a bipartite system with entanglement entropy $S$, the bond dimension must exceed $e^S$ in order to accurately describe the system.
As a special case, if $S=0$, then $s_l=\delta_{l,0}$, and the required bond dimension is 1.
Thus, for systems exhibiting strong entanglement, a large bond dimension is required to achieve an accurate simulation.
The choice of tensor network structure, particularly the MPS ordering or more generally the TTNS tree structure, affects the bipartition of the system and consequently the entanglement entropy. 
Structures that minimize bipartite entanglement entropy are preferred, as they allow for more efficient simulations. Finding the globally optimal structure for a given Hamiltonian is believed to be a challenging problem. 
In practice, tensor network structures are often designed based on heuristics or the nature of the system's interactions.

\subsection{Tensor Network Structures}
\label{sec:tns}
\begin{figure}[ht]
    \centering    \includegraphics[width=.85\textwidth]{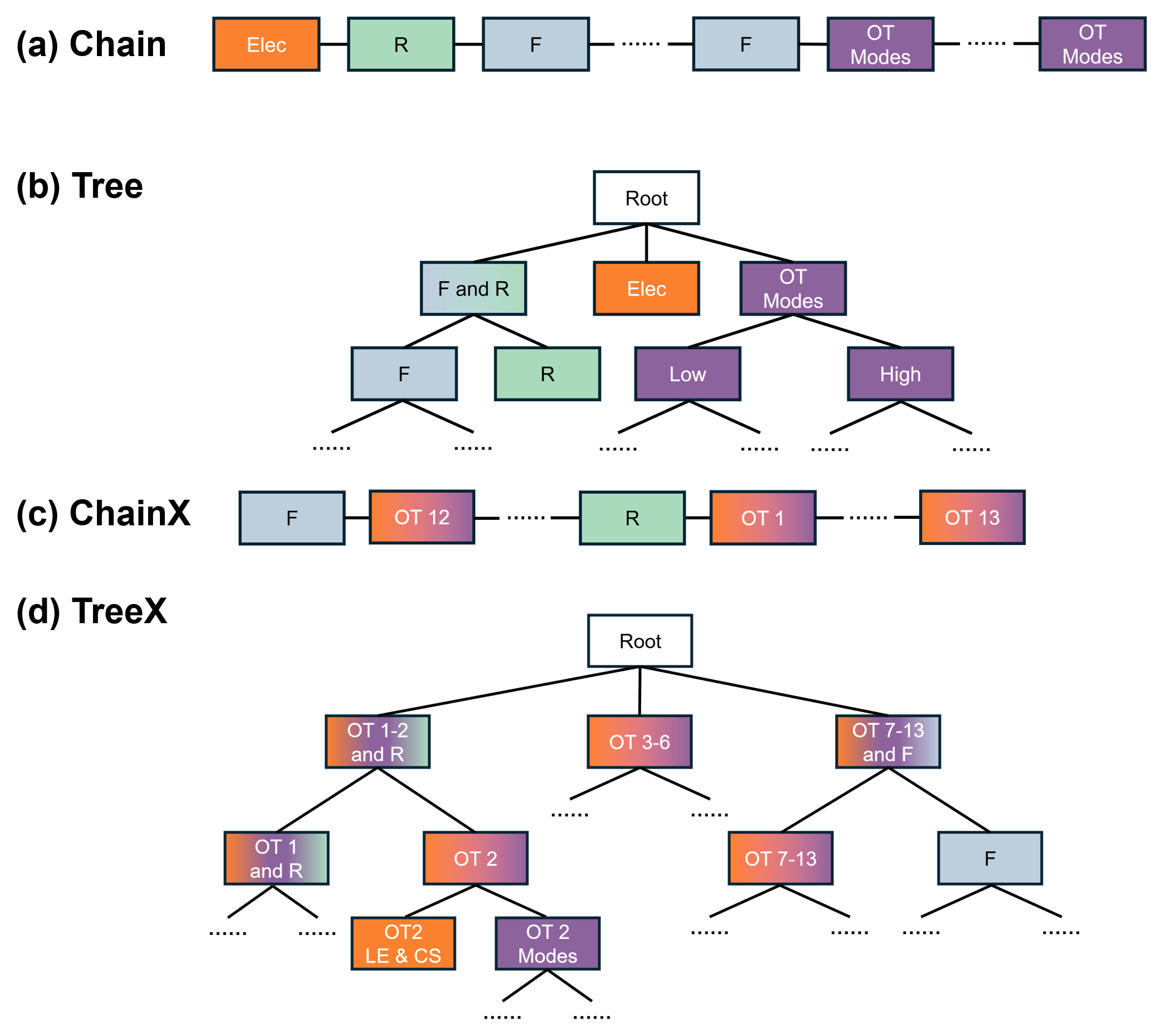}
    \caption{The tensor network topologies used in this work: (a) \rev{A MPS linear chain structure following the reference work}; (b) A TTNS structure following the reference work; (c) and (d) are additional MPS and TTNS structure proposed in this work. The four structures are denoted as  ``Chain'', ``Tree'', ``ChainX'' and ``TreeX'' respectively. The physical indices are ommited. The colors indicate different types of degrees of freedom.}
    \label{fig:diagram}
\end{figure}

In this study, we explore \rev{four} different tensor network structures, which are illustrated in Fig.~\ref{fig:diagram}.
The first is a standard MPS, shown in Fig.~\ref{fig:diagram}(a), which takes a linear form. 
The first site of the MPS contains all electronic states, including 13 LE states and 13 CS states.
This is followed by the $R$ site, and then 8 $F$ vibration sites.
Finally, the local OT vibrations are appended molecule by molecule.
The site ordering of the MPS is the same as the site ordering in the reference work~\cite{dorfner2024comparison}.

The second structure is a tree tensor network based on the reference work.
As shown in Fig.~\ref{fig:diagram}(b), the root node has 3 children.
The first child contains all 8 $F$ modes and the $R$ mode.
The second child contains the 26 different electronic states.
The third child contains all local OT vibrational modes.
The OT vibrational modes are divided into two subtrees based on the vibration frequency: low frequency modes and high frequency modes.
High frequency modes are defined as the modes whose $\omega_{\textrm{OT}, l} > 300$ meV.
According to Table~\ref{tab:vibrations}, there are two high frequency modes per OT molecule.
\rev{The topology of this tree is identical to the one used in the reference work}.
In the tree structure, a few nodes that are close to the leaves in the OT subtree have three children.
Since in this work we use fixed uniform bond dimension across the whole tree,
such ternary nodes become too large and lead to an unnecessary computational bottleneck.
\rev{To address this, we reduce the bond dimension associated with these nodes and their children to $\frac{1}{8}$ of the bond dimension of the entire tree.
Since these nodes are close to the leaves, this reduction in bond dimension has minimal impact on accuracy.
The bond dimension is the only difference between Tree and the TTNS structure employed in the reference work.
}
\rev{
We also note that although the root node in Fig.~\ref{fig:diagram}(b) and Fig~\ref{fig:diagram}(d) has three children, 
it is not the bottleneck of the computation.
This is because as the root of the tree, the total number of virtual indices it carries is three, which is comparable to other nodes that have two children and one parent. Therefore, a root node with three children ensures balanced computational cost over all nodes in the tree.
}

\rev{Next, we describe the tensor network structures proposed in this work, which are designed based on the calculation result from the Tree structure.
The third structure, shown in Fig.~\ref{fig:diagram}(c), is another MPS with different configuration compared to  Fig.~\ref{fig:diagram}(a).
For clarity, in this paper, we use ``Chain'' and ``ChainX'' to denote the MPS configurations shown in Fig.~\ref{fig:diagram}(a) and Fig.~\ref{fig:diagram}(c), respectively.
In ChainX, the LE state, CS state and the vibrations of each OT molecule are firstly grouped together.
Then, the OT$_1$ unit is placed at the center of the the chain, with OT units of even indices on the left side and OT units of odd indices on the right side. The $R$ mode is placed next to the OT$_1$ unit and the F modes are positioned at the beginning of the chain.
The structure is designed so that the degrees of freedom with the strongest entanglement entropy are placed in the center of the chain.
}

The last structure is a tree tensor network, shown in Fig.~\ref{fig:diagram}(d).
We use ``Tree'' and ``TreeX'' to denote the tree structures shown in Fig.~\ref{fig:diagram}(b) and Fig.~\ref{fig:diagram}(d), respectively.
In ``TreeX'', the electronic states and vibration states for each OT molecule are first grouped to form a subtree. The subtree is exemplified by the OT$_2$ subtree shown in Fig~\ref{fig:diagram}(c).
\revrev{The corresponding electronic node for OT$_2$ contains three bases: the vacuum state, the LE$_2$ state, and the CS$_2$ state. Quantum number constraints are applied to restrict the wavefunction to the single-exciton manifold. For discussions on grouping versus distributing electronic states in tensor networks within the MPS context, we refer readers to existing literature~\cite{baiardi2019large, sheng2024td}.}
The overall tree is then constructed based on the 13 subtrees.
The root node of the TreeX structure has 3 children.
The first child contains OT$_1$ and OT$_2$ subtree, as well as the $R$ vibration.
The second child contains the OT subtrees 3 to 6.
The third child contains the rest of the subtrees, as well as the $F$ vibrations.
TreeX is designed to minimize the entanglement entropy at the top layer of the tree.

For all tensor structures, \rev{the} harmonic oscillator eigenbasis is employed for the primitive basis of the vibrations, unless otherwise stated.
Following the reference work~\cite{dorfner2024comparison},
the number of the oscillator states for a given vibrational mode is set to $(g/\omega)^2+3g/\omega+N_b$,
where $g$ is the maximum coupling constant across all types of couplings, and $N_b$ is an adjustable offset.
In this work we employ $N_b=18$, which according to the reference paper should be enough for converged result~\cite{dorfner2024comparison}.
In \ref{sec:reproduce}, we show that using the discrete variable representation (DVR) basis does not significantly affect the calculated dynamics.

\revrev{The multiset formulation employs multiple, independent tensor network states, which are then linearly combined to represent the total wavefunction~\cite{fang1994multiconfiguration,kloss2019multiset}. This approach} has demonstrated good performance for the exciton dissociation model studied in this work~\cite{lindoy2025pyttn}. The multiset wavefunction allows different electronic states to couple with entirely different nuclear wavefunctions, which reduces the bond dimension for each individual tensor network state.
In this work, we focus on singleset tensor network states \revrev{where all electronic and vibrational states are encoded in one tensor network}. 
However, we are actively developing a multiset implementation, and a systematic benchmark comparing multiset and singleset tensor network states will be the focus of our future research.

Another important aspect of MPS and TTNS algorithm is the corresponding operators,
namely matrix product operators (MPO) and tree tensor network operators (TTNO)~\cite{schollwock2011density, chan2016matrix, szalay2015tensor}.
Similar to MPS and TTNS, MPO and TTNO are tensor network representations of quantum operators.
\rev{
Within ML-MCTDH, the Multi-Layer operators introduced by ML-Potfit are closedly related to TTNOs~\cite{otto2014multi}. 
ML-Potfit itself is a generalization of the Potfit algorithm~\cite{jackle1996product}, adapted for the multi-layer framework. 
Its primary function is to numerically construct the potential energy operator by fitting a continuous analytical potential energy surface into the required ML operator form.
}
In this work, MPO and TTNO are constructed exactly given the sum-of-product form of the operator.
This distinguishes our approach from standard ML-MCTDH, where the direct sum-of-product form is generally used for model systems.
In our previous work, we demonstrated that employing MPO provides a computational scaling advantage over the sum-of-product approach~\cite{ren2022time}.
We utilize the automatic MPO/TTNO construction algorithm based on bipartite graph theory~\cite{ren2020general, li2024optimal},
which efficiently generates the \revrev{most compact} MPO/TTNO with negligible computational cost.
Here, ``\revrev{most compact}'' means the exact MPO and TTNO with minimal bond dimension,
which will directly affect the computational cost.
The maximum MPO/TTNO bond dimension for Chain, Tree, ChainX and TreeX are 29, 29, 29, 14, respectively.
\rev{
For comparison, the corresponding Hamiltonian operator contains 503 terms in the sum-of-product form.
}
For both Chain and Tree structures, the maximum MPO/TTNO bond dimension is associated with the electronic node, because all 26 electronic states are represented in a single leaf node.
\revrev{
The 29 operaters include 26 diagonal operators for each electronic state, one identity operator to be coupled with  $\hat H^{\textrm{vib}}$, one electronic Hamiltonian operator $\hat H^{\textrm{e}}$ to be coupled with the identity operator of the vibrations, and 
one off-diagonal operator $\ket{\textrm{LE}_1} \bra{\textrm{CS}_1} + \textrm{H.C.}$ to be coupled with the $R$ mode.
In TreeX, the maximum bond dimension is reduced to 14 because the diagonal electronic operators are coupled with the vibrations and merged with $\hat H^{\textrm{e}}$ in the subtrees.
This is not possible in ChainX, where all diagonal electronic operators must individually couple with the $F$ mode placed at the boundary of the chain.
For more details of MPO/TTNO construction, we refer readers to our prior works, where we provide the specific matrix elements of MPO and TTNO for several typical models~\cite{ren2018time, li2024optimal}}.

The time evolution of MPS/TTNS is carried out using \rev{the TDVP-PS algorithm, which represents another key difference between our approach and standard ML-MCTDH.}
\revrev{The TDVP-PS algorithm for TTNS (PSI-ML-MCTDH) and its variants have been
   known for over a decade~\cite{lubich2015mctdh, lindoy2021time, lindoy2021time2, weike2021symmetries}, however, the variable mean-field (VMF)
   method remains predominant for time integration in the ML-MCTDH community.
   The VMF method is also employed in the reference work.
The PS algorithm features a sequential sweep over the tensor network to perform the time evolution, while VMF
updates all nodes in the tensor network synchronously.}
We shall discuss the difference between the TDVP-PS and the VMF in \ref{sec:reproduce}.
In this work, unless otherwise stated, the 1-site TDVP-PS algorithm is employed.
The product between a matrix exponential and a vector is solved by the Krylov solver, or more specifically the short iterative Lanczos algorithm for the Hermitian system studied here.
For all results reported, the time evolution step is set to 1 fs. 
During initial benchmarks, we found that TDVP-PS supports \revrev{a} much longer time step than 1 fs, 
yet for the purposes of this paper, we choose 1 fs as the time step for denser data points.
\rev{Benchmark data for the time step is provided in \ref{sec:convergence}.}

\section{Results and Discussion}
\label{sec:results}
\begin{figure}[ht]
    \centering    \includegraphics[width=\textwidth]{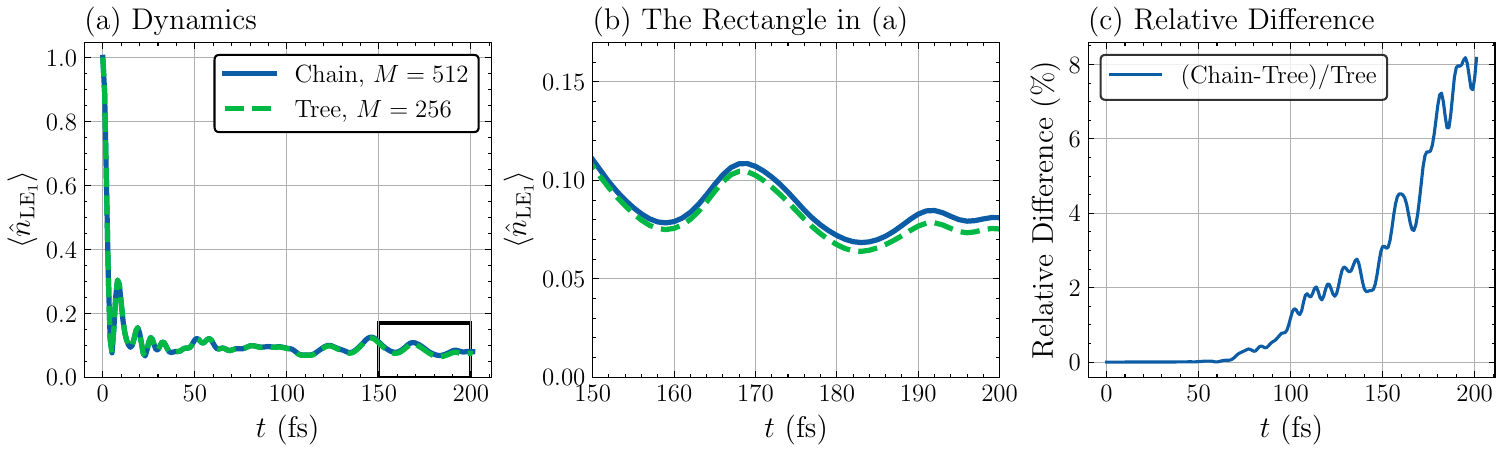}
    \caption{The difference in the $\braket{\hat n_{\textrm{LE}_1}}$ dynamics calculated by Chain and Tree.
    (a) The full dynamics from 0 to 200 fs. (b) The dynamics at the long time limit, i.e., from $150$ to $200$ fs.
    (c) The relative difference between the $\braket{\hat n_{\textrm{LE}_1}}$ values, using Tree values as the reference.}
    \label{fig:dynamics-raw}
\end{figure}

We first compare the dynamics of $\braket{\hat n_{\textrm{LE}_1}}$ by Chain and Tree, calculated by \textsc{Renormalizer}, in Fig.~\ref{fig:dynamics-raw}.
The employed bond dimensions $M$ for Chain and Tree are 512 and 256, respectively.
The full dynamics in Fig.~\ref{fig:dynamics-raw}(a) agrees with the reference paper~\cite{dorfner2024comparison},
which describes the depopulation of the LE state to other LE and CS states.
Moreover, the difference between the Chain and Tree results is almost negligible.
In Fig.~\ref{fig:dynamics-raw}(b) we zoom in on region marked by the rectangle in Fig.~\ref{fig:dynamics-raw}(a)
for a clear visualization of the difference at the long time limit.
From Fig.~\ref{fig:dynamics-raw}(b), we estimate that  the difference between the Chain and Tree results is at the order of 0.005.
In Fig.~\ref{fig:dynamics-raw}(c), we illustrate the relative difference of $\braket{\hat n_{\textrm{LE}_1}}$, using the Tree results as the reference.
In general, the relative difference increases over time, and the maximum difference is approximately 8\%.
We note that the same MPS ordering and TTNS structure as in the reference work are employed to produce Fig.~\ref{fig:dynamics-raw}.
Overall, Fig.~\ref{fig:dynamics-raw} shows that Chain and Tree produce consistent dynamics, with the differences being negligible for most practical purposes.

\begin{figure}[t]
    \centering    \includegraphics[width=.9\textwidth]{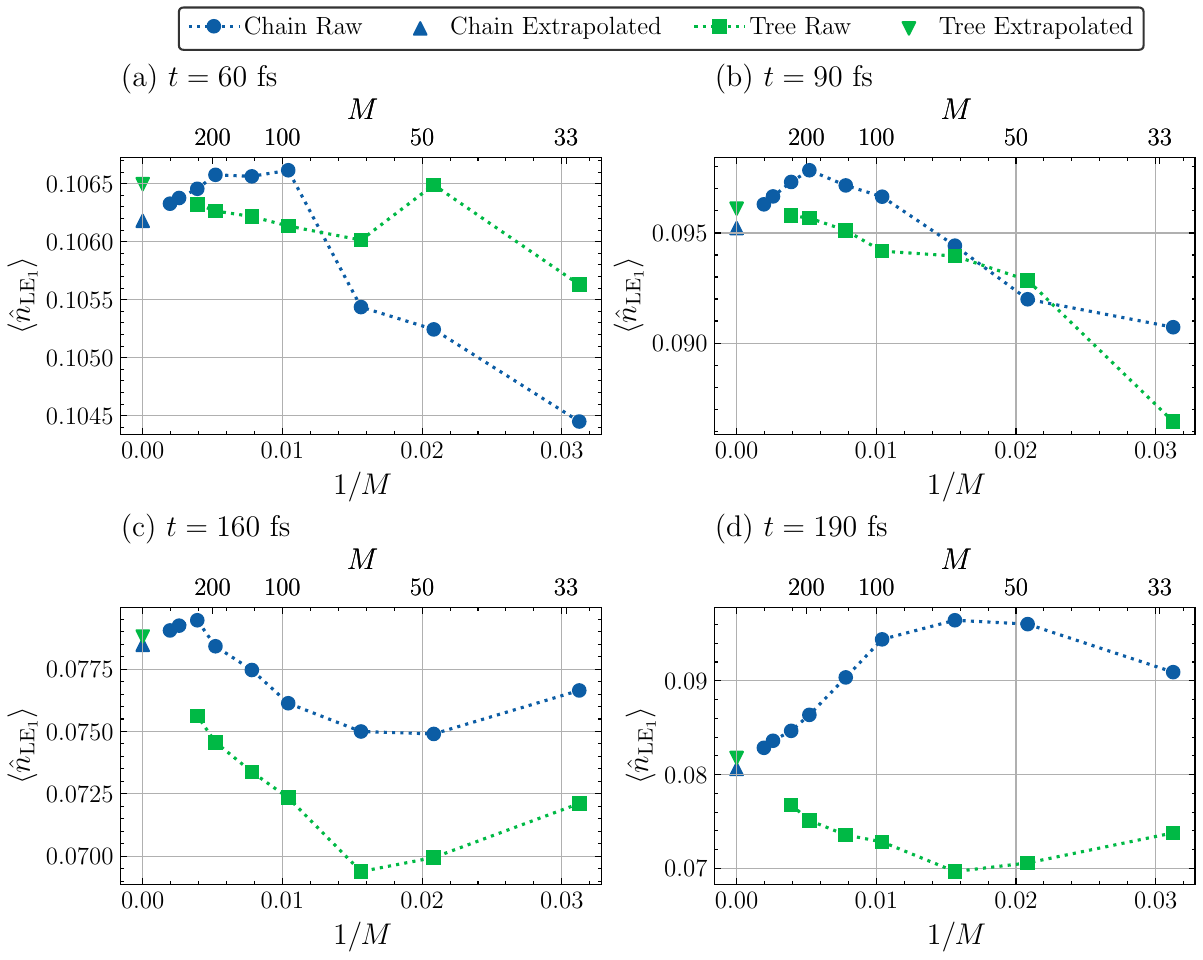}
    \caption{Convergence of $\textrm{LE}_1$ state occupation with bond dimension.
    The occupation is plotted against inversed bond dimensionas $1/M$ at (a) $t=60$ fs, (b) $t=90$ fs, (c) $t=160$ fs and (d) $t=190$ fs. The expected values at $1/M=0$ are obtained from linear fits to the two smallest $1/M$ data points.}
    \label{fig:ext}
\end{figure}

We then provide further evidence that in our calculation Chain and Tree are converging to the exact limit.
In Fig.~\ref{fig:ext} we show how $\braket{\hat n_{\textrm{LE}_1}}$ evolves when $M\rightarrow \infty$.
More specifically, we choose $t=60, 90, 160$ and 190 fs as several \rev{representative} time frames, and plot 
$\braket{\hat n_{\textrm{LE}_1}}$ against $1/M$.
For reference, we include the full dynamics calculated with different $M$ in Fig.~\ref{fig:convergence} in \ref{sec:convergence}.
In Fig.~\ref{fig:ext}, we observe that as $M$ increases and $1/M$ decreases, the results from Chain and Tree converge.
We then estimate the value of $\braket{\hat n_{\textrm{LE}_1}}$ in the $1/M\rightarrow 0$ limit by linear extrapolation.
The technique is inspired by the extrapolation scheme employed in large scale static DMRG calculations~\cite{white2005density, li2019electronic, larsson2025benchmarking}.
To exclude the noise induced by the calculations with small $M$, we employ only the two data points with the smallest $1/M$ for the extrapolation.
Fig.~\ref{fig:ext} shows that
this simple extrapolation scheme reduces the difference by Chain and Tree when $t=160$ and 190 fs.
We note that such extrapolation is only valid when the bond dimension is sufficiently large.
Fig.~\ref{fig:ext} shows that the dependence of $\braket{\hat n_{\textrm{LE}_1}}$ on $1/M$ is highly nonlinear.
As a result, extrapolation using the datapoint with $M<100$ \rev{(for Tree)} or even $M<200$ \rev{(for Chain)} will likely result in even larger errors compared to the raw data.
\revrev{This behavior is fundamentally different from static DMRG calculations. There, the energy is directly minimized, resulting in a monotonic relationship with $1/M$ that enables reliable extrapolation.}
\rev{
Empirically, we found that the extrapolation improves the consistency of results across different tensor network structures particularly at the long time limit. Therefore, we consider the extrapolated results to be more reliable at the long time limit. However, in other computational tasks where such extensive benchmarks are unavailable, the extrapolation should be applied with caution.}

Having confirmed that the dynamics in Fig.~\ref{fig:dynamics-raw} is numerically exact, we investigate the origin of the discrepancy reported in the reference work.
Following the prescription described in the reference paper, we reproduced these results using \textsc{Renormalizer}, which are shown in Fig.~\ref{fig:reproduce} in \ref{sec:reproduce}.
We find that the bond dimensions of both Chain and Tree are the key to the difference.
On the one hand, the bond dimension of Chain is expanded using 2-site TDVP-PS with singular value truncation,
and the bond dimension becomes fixed after the maximum bond dimension reaches a target value.
While effective for short-time dynamics, this method creates a bond dimension distribution biased toward early-time entanglement.
This \rev{becomes the source of the error} for the studied model in which the entanglement spreads from the OT$_1$ molecule to other molecules over time.
In our approach for the results in Fig.~\ref{fig:dynamics-raw}, a fixed uniform bond dimension across the whole MPS chain is employed, avoiding the problem.
On the other hand, in the reference work, the maximum bond dimension of the Tree calculation is $M=40$.
However, our convergence analysis in Fig.~\ref{fig:ext} shows that \rev{further increasing $M$ can enhance the long-time accuracy}.
As a result, both the Chain and Tree results deviate from the exact solution \rev{by a small yet non-negligible margin}.
Further details are provided in \ref{sec:reproduce}.
\rev{
We note that despite the relatively small bond dimension employed in the reference work,
the short-time dynamics agree nearly perfectly. 
Additionally, the maximum absolute difference in \revrev{electronic state occupancy} at the long-time limit is only 0.05, which is often sufficient for many practical applications.
}

\begin{figure}[h]
    \centering    \includegraphics[width=.6\textwidth]{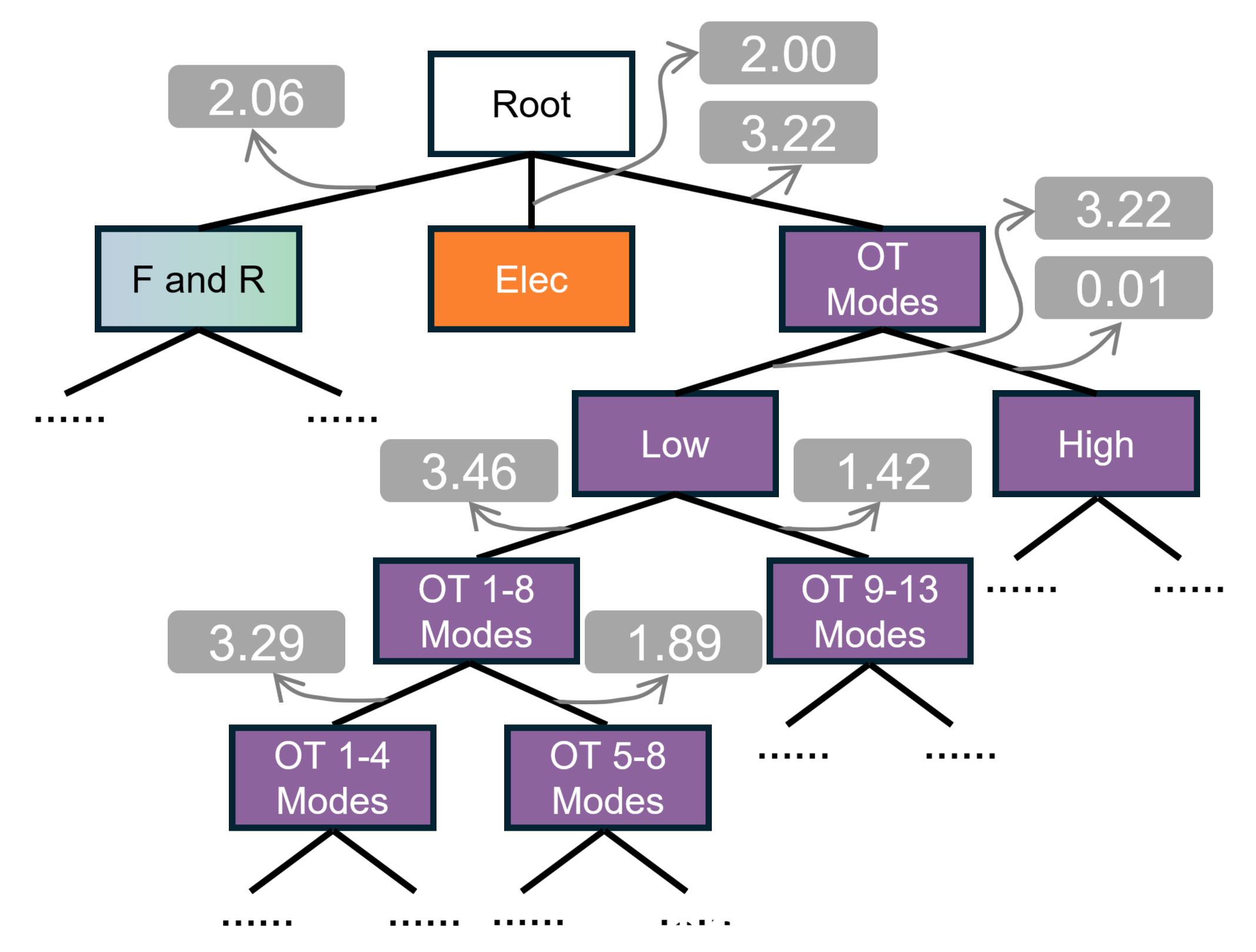}
    \caption{The von Neumann entanglement entropy distribution of Tree when $t=200$ fs.}
    \label{fig:entanglement}
\end{figure}

\rev{The residual difference between Chain and Tree structures with large bond dimension} motivated the development of optimized tensor network structures.
\rev{For MPS and TTNS, the new structures are termed ChainX and TreeX, respectively, and are shown in Fig.~\ref{fig:diagram} (c) and (d).}
\rev{ChainX and TreeX are} designed based on the \rev{entanglement} analysis of Tree calculations, shown in Fig.~\ref{fig:entanglement}.
The bonds with \rev{the} largest entanglement entropy are shown as they are the bottleneck of the computation.
Fig.~\ref{fig:entanglement} reveals that the low frequency OT modes  exhibit significantly stronger entanglement
compared to the $F/R$ vibrations, the electronic states, and the high frequency modes.
High-frequency modes show negligible entanglement, allowing flexible placement in the network without \rev{affecting} the calculated dynamics.
Furthermore, the low frequency modes associated with the smallest OT index have the strongest entanglement.
Guided by these observations, TreeX groups OT 1-2, OT 3-6 and OT 7-13 as the top subtrees, to distribute the entanglement evenly.
\rev{ChainX is inspired by TreeX, where the OT 1 unit with the strongest entanglement is placed in the middle of the chain.}
We note that although our numerical experiments confirm that TreeX reduces errors compared to other tensor network structures, it is unlikely the optimal tree structure for this model and further \rev{structural} optimization could lead to additional enhancements~\cite{larsson2019computing, li2022fly}.

\begin{figure}[t]
    \centering    \includegraphics[width=.7\textwidth]{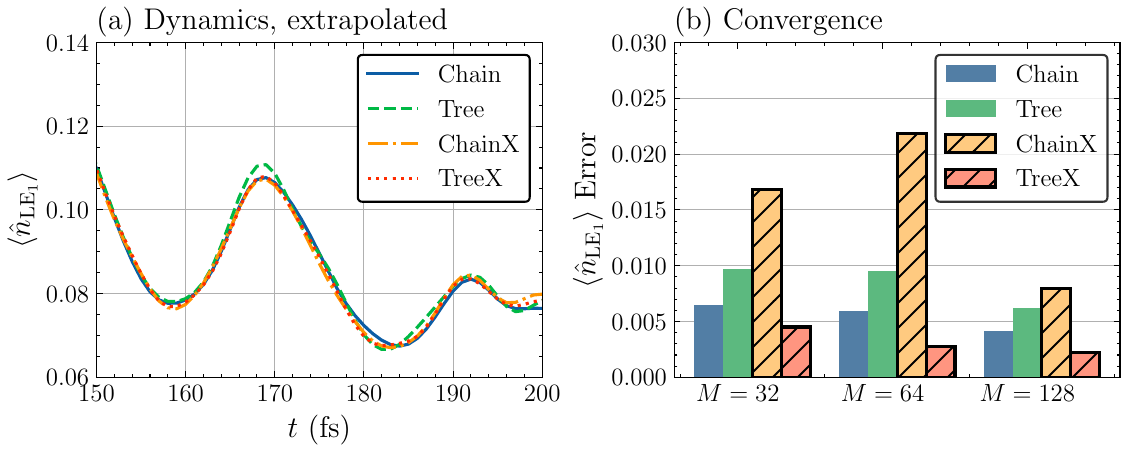}
    \caption{\rev{The performance of ChainX and TreeX structure.  (a) The extrapolated $\braket{\hat n_{\textrm{LE}_1}}$  dynamics  from Chain, Tree, ChainX and TreeX calculations; (b) The convergence behavior with respect to the bond dimension $M$ for Chain, Tree, ChainX and TreeX.}
    The error in (a) is measured by the time-average deviations of $\braket{\hat n_{\textrm{LE}_1}}$ from $t=150$ to 200 fs from the extrapolated $\braket{\hat n_{\textrm{LE}_1}}$ reference values.
}
    \label{fig:tree4}
\end{figure}

Our \rev{ChainX/TreeX} calculations, performed with bond dimensions up to \rev{$M=512$ and $M=256$ respectively} and combined with the extrapolation scheme from  Fig.~\ref{fig:ext}, are shown in Fig.~\ref{fig:tree4}(a).
\rev{
Chain, Tree, ChainX and TreeX results demonstrate exceptional agreement.
The corresponding maximum absolute difference is estimated to be approximately 0.002, with an average error of approximately 0.001.}
Such precision is remarkable for the complex quantum system and highlights the reliability of numerically exact tensor network methods.
The complete set of extrapolated results for all 26 electronic states is provided in \ref{sec:all-dof}, and all source data are available in our repository~\cite{data}.
We believe the data will serve as accurate benchmark data for the further development of quantum dynamics algorithms.

In Fig.~\ref{fig:tree4}(b) we show that TreeX shows the fastest convergence compared to Chain, Tree and ChainX structures.
Using the extrapolated values from Fig.~\ref{fig:tree4}(a) as reference,
we quantify the error as the time-averaged absolute deviation between $t=150$ and 200 fs:
\begin{equation}
\label{eq:error}
    \textrm{Error} = \frac{\sum_{t=150}^{200}\abs{n(t) - n_\textrm{ref}(t)}}{200-150} \ .
\end{equation}
Here $n$ represents the occupation of the LE$_1$ state.
\rev{The reason for choosing the 150-200 fs window is two-fold. 
The first is that the error in  this time window dominates the total error. 
The second is that throughout the manuscript we often focus on the dynamics in this time range for detailed analysis.
}
Our analysis reveals that TreeX achieves comparable accuracy at $M=32$ to what Chain attains at $M=128$.
And when $M=128$ the average error of TreeX is already as small as 0.002, approaching the uncertainty limit of our reference values.
This enhanced performance is consistent with the convergence trends shown in Fig.~\ref{fig:convergence},
and demonstrates that the optimized tree structure more effectively captures the \rev{entanglement} pattern of the system.
Meanwhile, ChainX exhibits the largest error compared with the other three tensor network structures. This is because, although the units with the strongest entanglement are placed in the middle, the ordering of the sites in ChainX does not follow the original site ordering of the model. This misalignment typically leads to large entanglement entropy and consequently a large error in MPS structures.
\revrev{In comparison, TreeX is able to be mapped onto a one dimensional topology while preserving site ordering
and more critically to distribute entanglement entropy evenly across subtrees. ChainX cannot achieve this balanced distribution simply by placing the most entangled units in the middle.}
The contrast of the different convergence rates highlights the importance of careful structural design for achieving optimal computational efficiency.

\begin{figure}[t]
    \centering    \includegraphics[width=.7\textwidth]{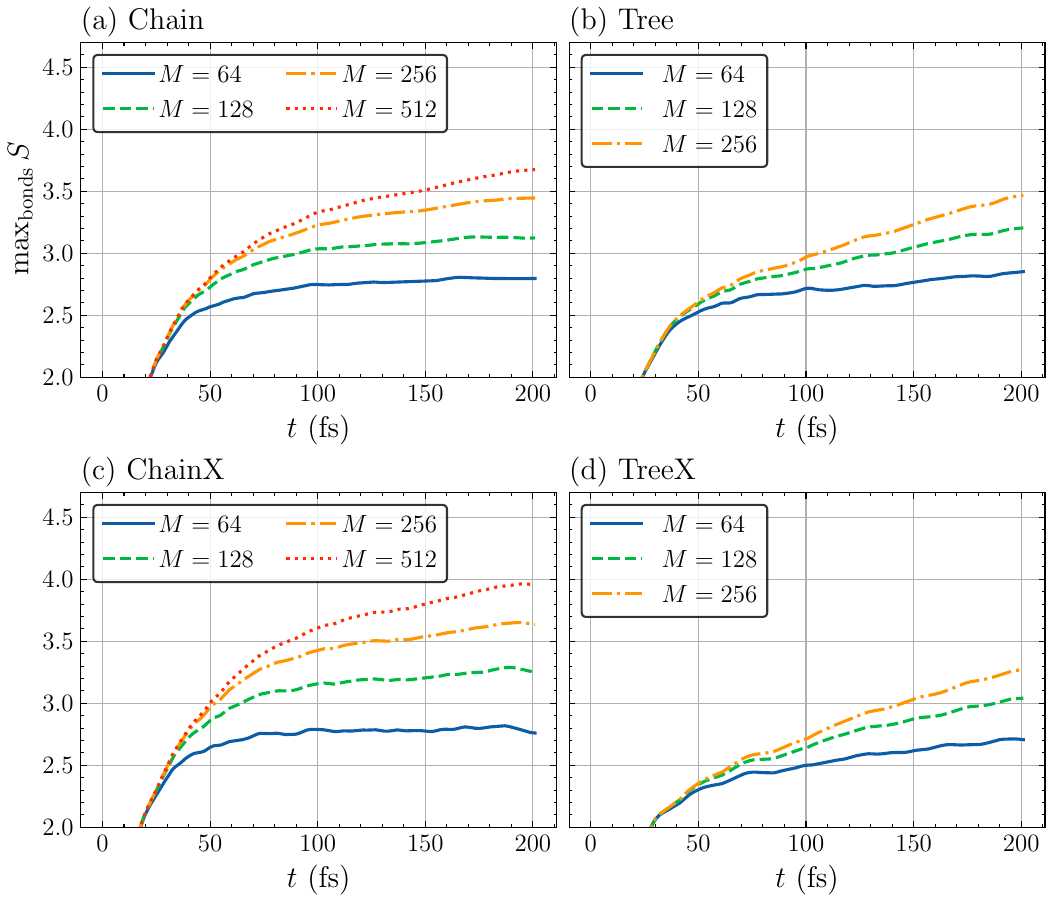}
    \caption{The time evolution of the maximum von Neumann entanglement entropy for  (a) Chain, (b) Tree (c) ChainX and (d) TreeX structures.
    The entropy values represent the maxima across all virtual bonds in each network architecture.}
    \label{fig:entanglement3}
\end{figure}

In order to gain deeper insight into the different convergence rate of Chain, Tree, \rev{ChainX} and TreeX, in Fig.~\ref{fig:entanglement3} we analyze the time evolution of the von Neumann entanglement entropy $S$
for the \rev{four} methods.
The maximum $S$ across all virtual bonds in the correponding  tensor network structure is reported.
Since in this work we have employed a uniform bond dimension distribution, the maximum $S$ determines the error of the calculation.
\rev{As shown in Eq.~\ref{eq:compression-error}, the tensor network compression error at a particular bond is the sum of the square of the discarded singular values.
Thus, the bond with the largest $S$ will have the largest absolute compression error when a fixed bond dimension is applied uniformly. In other words, this single largest error becomes the dominant and limiting source of inaccuracy for the entire calculation and determines its overall error.
}

The spatial distribution of maximum $S$ in the tensor network varies between methods.
For Chain, the maximum $S$ appears between the local vibrations of OT$_1$ and OT$_3$ molecules, which is shown in Fig.~\ref{fig:entanglement-m}.
For Tree, the maximum $S$ is associated with the bond that connects the local vibrations modes of OT$_1$ to OT$_7$ to the rest of the system, which is show in Fig.~\ref{fig:entanglement}.
\rev{For ChainX, the maximum $S$ appears between the $R$ mode and the OT$_1$ unit, which aligns with the design principle.}
For TreeX, the maximum $S$ is designed to appear in the top layer.

In Fig.~\ref{fig:entanglement3}, we first find that the convergence of maximum $S$ is much slower than the convergence of the occupation for all methods.
Thus, the maximum $S$ may serve as a strict criteria for numerical convergence.
Additionally, while Tree initially shows slower entanglement growth than Chain, its entanglement increases rapidly thereafter, ultimately reaching comparable values by $t = 200$ fs.
\rev{ChainX shows the highest entanglement entropy, which leads to the larger error observed in Fig.~\ref{fig:tree4}(b).}
In contrast, TreeX shows the slowest entanglement growth.
As a result, when $t=200$ fs, TreeX has the smallest $\max_{\rm{bonds}}S$ across the \rev{four} methods.
This suppressed entanglement accumulation directly correlates with TreeX's superior accuracy shown in Fig.~\ref{fig:tree4}(b).

\begin{figure}[t]
    \centering    \includegraphics[width=.5\textwidth]{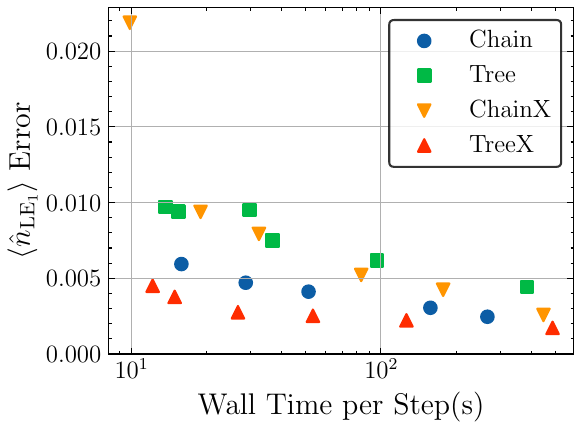}
    \caption{The average error \rev{defined in Eq.~\ref{eq:error}} versus the wall time per step for Chain, Tree, ChainX and TreeX.}
    \label{fig:wall_time}
\end{figure}

Lastly, we discuss the efficiency of the algorithms.
\rev{For the same bond dimension $M$, the MPS structures require less memory,
which allows us to perform $M=512$ calculation with Chain/ChainX}.
Both Tree and TreeX have similar memory requirement.
However, the TTNO of Tree has a larger bond dimension, leading to larger intermediate tensors and a higher computational cost compared to TreeX.
MPS calculation with $M=256$ and $M=512$ and TreeX calculations with $M=256$ are carried out on a V100 (32GB) GPU card in combination with 4 cores of an AMD EPYC 74F3 CPU~\cite{li2020numerical}.
Tree calculation with $M=256$ is carried out on a A100 (80GB) GPU card in combination with Intel Xeon Gold 6226R CPU, due to its higher memory consumption.
\rev{
The thermal design power (TDP) for the AMD EPYC 74F3 CPU and the NVIDIA GPUs is rated at around 300 W. 
The TDP value represents their maximum thermal output under full load. Our calculations account for a more typical usage scenario where only one-sixth of the CPU cores are active, and GPU utilization averages 50\%. Based on this reduced load, we estimate the total power consumption for this workload to be approximately 200 W.
}
Fig.~\ref{fig:wall_time} presents the tradeoff between the average error and computational efficiency for Chain, Tree, ChainX and TreeX.
The wall times reported here include the time evolution and the calculation of all physical observables, such as the bond singular values and the RDMs of all degrees of freedom in the system.
As shown, the TreeX structure demonstrates the best balance between accuracy and computational cost, achieving the smallest average error while requiring less wall time.
\rev{Chain is more efficient than Tree, because of its lower computational scaling}. This figure highlights that tree structure design is crucial to the success of the tree tensor network algorithms.

\section{Conclusion and Outlook}
In this work, we \rev{develop from a previous work and perform} systematic benchmark of \rev{MPS and TTNS} for a challenging exciton dissociation model.
\revrev{Using the TDVP-PS algorithm, where MPS time evolution corresponds to TD-DMRG and TTNS to PSI-ML-MCTDH,}
we demonstrate \revrev{exceptional agreement between MPS and TTNS, with an average difference in electronic state occupancy of only 0.001}. This is achieved \revrev{through}: (1) the use of sufficiently large bond dimensions ($M=512$ for MPS and $M=256$ for TTNS), which alone reduces the difference to \revrev{$\sim 0.005$}; (2) the development of an optimized tree structure termed TreeX that shows superior convergence properties.
\rev{The exceptional efficiency of TreeX is demonstrated by its achievement of an average \revrev{occupancy} error $< 0.005$ with only $M=32$ bases}.
\rev{We note that TreeX is not designed a priori, but rather based on the calculated entanglement entropy from TTNS simulations. While we also designed a similar new MPS structure termed ChainX, the error was higher in comparison. Our results show that the structural flexibility of the tree, compared to a linear topology, offers more flexibility to reduce entanglement entropy and enhance computational efficiency.}

Our calculation confirms that both \rev{MPS and TTNS  are numerically-exact ansätze}
that are capable of achieving high accuracy for the simulation of complex chemical systems with moderate cost.
We hope the high-accuracy benchmark data presented here will facilitate future developments of not only tensor network algorithms but also other types of quantum dynamics methods.
\rev{At the same time, our study highlights practical challenges in tensor network simulations. High-quality results require extensive convergence tests for parameters such as bond dimension and tree structures. To overcome this, we need new descriptors to quantify errors in time-evolved observables, as well as algorithms that can efficiently and reliably generate near-optimal tensor network structures without extensive trial calculations.}
We expect continued cross-fertilization between TD-DMRG and ML-MCTDH will enable the simulation of even more challenging quantum systems with unprecedented accuracy and efficiency.

\appendix

\renewcommand{\thesection}{Appendix \Alph{section}} 
\counterwithin{figure}{section} 
\counterwithin{table}{section}  

\renewcommand{\thefigure}{\Alph{section}\arabic{figure}}
\renewcommand{\thetable}{\Alph{section}\arabic{table}}

\section{Model Parameters}
\label{sec:params}
In  this section, we list the specific model parameters for the exciton dissociation model studied in this work.
All parameters are directly adopted from the reference work~\cite{dorfner2024comparison}.
In Table~\ref{tab:epsilon_cs} we list the on-site energy of the 13 CS states,
used in Eq.~\ref{eq:ham-e}.
In Table~\ref{tab:vibrations} we list the frequencies and coupling constants of the $F$ modes and  the OT local vibrational modes.

\begin{table}[htbp]
    \centering
    \begin{tabular}{cc}
        \hline
        $n$ & $\epsilon_n^{\textrm{CS}}$ (meV) \\
        \hline
        1 & 0.0 \\
        2 & 33.6 \\
        3 & 47.4 \\
        4 & 56.0 \\
        5 & 61.8 \\
        6 & 65.7 \\
        7 & 68.4 \\
        8 & 70.0 \\
        9 & 70.9 \\
        10 & 71.2 \\
        11 & 71.1 \\
        12 & 70.5 \\
        13 & 69.5 \\
        \hline
    \end{tabular}
    \caption{The on-site energy of the CS states.}
    \label{tab:epsilon_cs}
\end{table}

\begin{table}[htbp]
    \centering
    \begin{tabular}{cccccc}
    \hline
    $l$ & $\omega_{F,l}$ & $\omega_{\textrm{OT}, l}$ & $g_{F, l}$ & $g_{\textrm{OT}, l}$ & $g'_{\textrm{OT}, l}$ \\ \hline
    1 & 200.025 & 401.283 & 45.246 & 7.017 & 4.035 \\ 
    2 & 184.269 & 397.773 & 65.701 & -0.077 & 2.921 \\ 
    3 & 177.853 & 182.714 & -40.280 & -67.849 & -129.712 \\ 
    4 & 141.11 & 178.531 & -17.511 & 57.668 & 46.885 \\ 
    5 & 93.952 & 134.550 & 28.026 & -40.145 & -32.908 \\ 
    6 & 79.933 & 111.848 & -13.629 & 11.68 & 36.591 \\ 
    7 & 55.892 & 42.621 & -23.732 & -10.784 & -20.211 \\ 
    8 & 33.264 & 18.316 & 9.86 & -12.309 & -7.77 \\ \hline
    \end{tabular}
    \caption{The frequencies and coupling constants of the $F$ modes and the OT local vibrational modes. \rev{The units are meV.}}
    \label{tab:vibrations}
\end{table}

\section{Convergence Benchmark}
\label{sec:convergence}
\begin{figure}[htbp]
    \centering    \includegraphics[width=\textwidth]{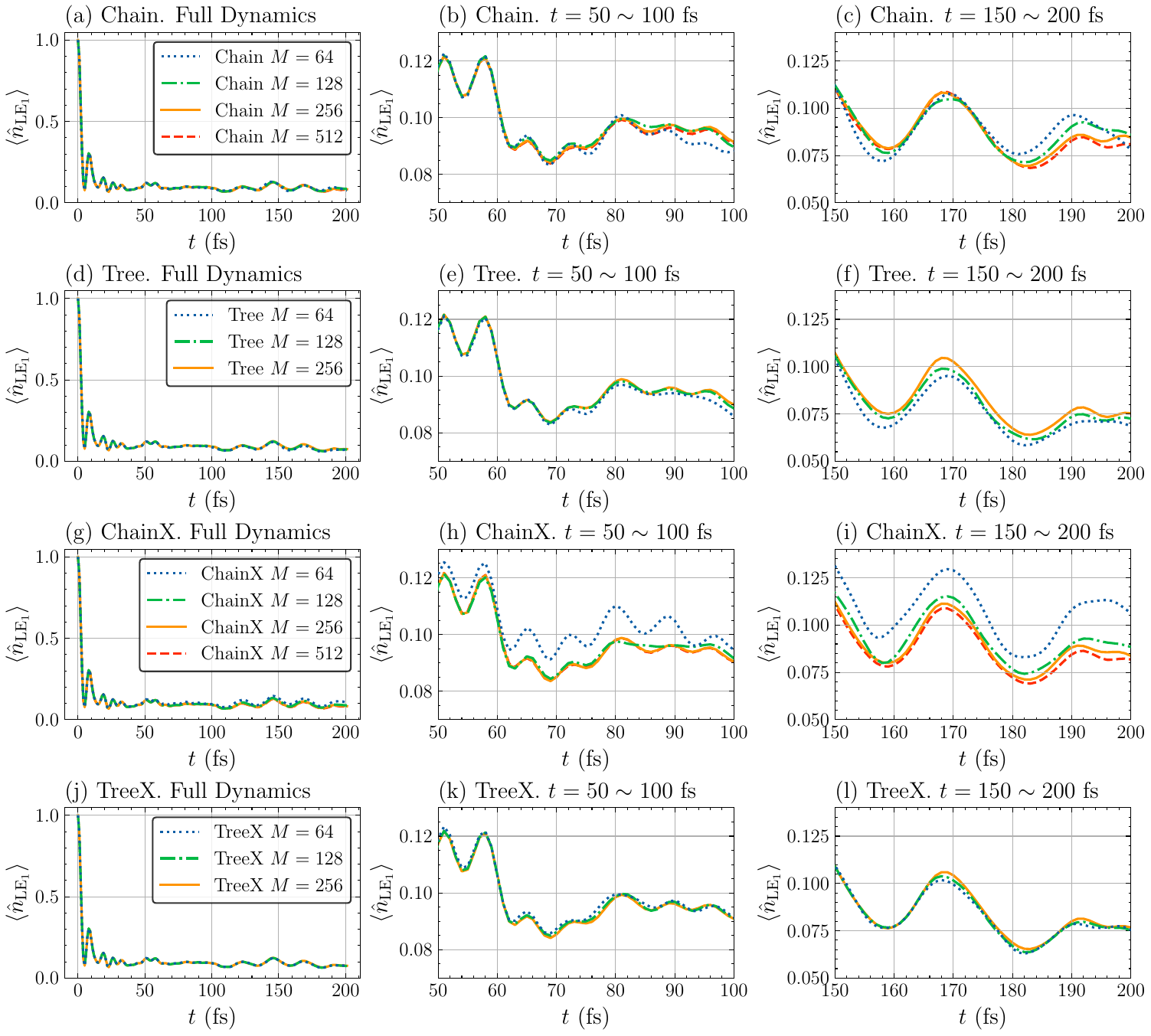}
    \caption{The convergence analysis of (a-c) Chain, (d-f) Tree, \rev{(g-i) ChainX and (j-l) TreeX}.
    In the first, second and third column the full dynamics, the dynamics from 50 to 100 fs, and the dynamics from 150 to 200 fs are shown, respectively.}
    \label{fig:convergence}
\end{figure}

In this section, we present a convergence analysis of \rev{both the bond dimension and the time step}.
The dynamics of $\braket{\hat n_{\textrm{LE}_1}}$ with different bond dimension $M$ is shown in Fig.~\ref{fig:convergence}.
While all \rev{four tensor network structures} appear converged at the full scale in the first column, closer inspection of specific time windows reveals differences. 
The $50\sim100$ fs region in the second column shows satisfactory convergence for large $M$ values across all methods. 
However, in  the long time limit ($t=150\sim 200$, third column), \rev{Chain, Tree and ChainX} exhibit significant convergence challenges, while TreeX maintains stable convergence behavior.
From Fig.~\ref{fig:convergence}(c) and Fig.~\ref{fig:convergence}(f),
we see that if $M$ is not sufficiently large, Chain consistenly overestimates the occupation,
while Tree tends to underestimate the occupation.
These opposing trends explain the discrepancies observed between the two methods 
when \rev{small} bond dimensions are used.
Nonetheless, Chain and Tree are converging to the same result as $M$ increases.
The convergence patterns support our extrapolation approach shown in Fig.~\ref{fig:ext}.

\begin{figure}[htbp]
    \centering    \includegraphics[width=\textwidth]{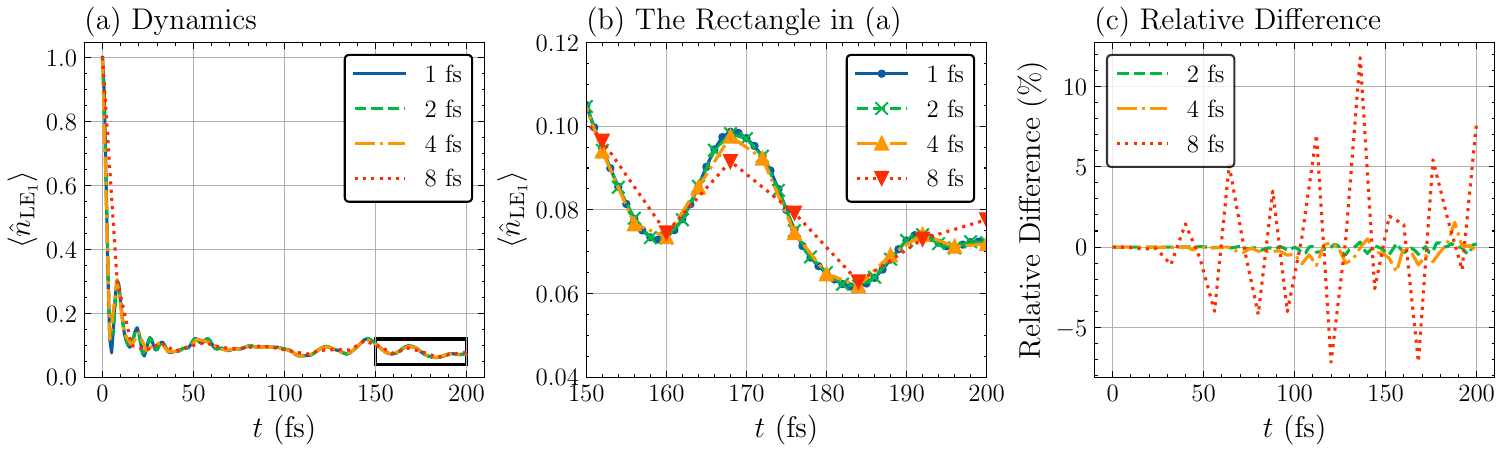}
    \caption{\rev{The difference in the $\braket{\hat n_{\textrm{LE}_1}}$ dynamics calculated by Tree with increasing time step.
    (a) The full dynamics from 0 to 200 fs. (b) The dynamics at the long time limit, i.e., from $150$ to $200$ fs.
    (c) The relative difference between the $\braket{\hat n_{\textrm{LE}_1}}$ values, using values obtained by a time step of 1 fs as the reference.}}
    \label{fig:dt}
\end{figure}

\rev{In Fig.~\ref{fig:dt} we show the effect of the time step on the dynamics obtained by Tree with $M=128$.
We find that using a time step of 4 fs will leads to an error smaller than the estimated uncertainty caused by insufficient bond dimension.
Thus for the benchmark data reported in this work in principle we can use 4 fs as the time step.
However, as shown in Fig.~\ref{fig:dt}(b), using such large time step will cause sparse data points and discontinuous dynamics. Thus in the main text we choose to employ 1 fs as the time step for denser data points.}

\rev{We note that the large time step does not cause large error despite the high frequency modes in the model.
This is because these high frequency modes shows neglegible entanglement entropy with the rest of the system, as shown in Fig.~\ref{fig:entanglement}. 
Therefore, one may take the mean-field picture to deal with these high frequency modes.
In other words, their fast motion constitudes a mean-field to the dynamics of the electron.
For the purpose of calculating the electron occupations, it's not necessary to exactly track the high-frequency nuclear motion.
On the hand hand, the projector-splitting integrator empoloyed in this work features sequential update of each node in the tensor network. The integrator is then able to adaptively perform the time evolution of each node with different number of steps (the number of Krylov vectors in matrix exponential).
The symplectic feature of the integrator also helps to ensure the numerical accuracy at long time scale.
In our previous work, we have found that TDVP-PS tolerant exceptionally long time step in MPS simulations~\cite{li2020numerical}.}

\section{Reproducing the Reference}
\label{sec:reproduce}
\begin{figure}[htbp]
    \centering    \includegraphics[width=\textwidth]{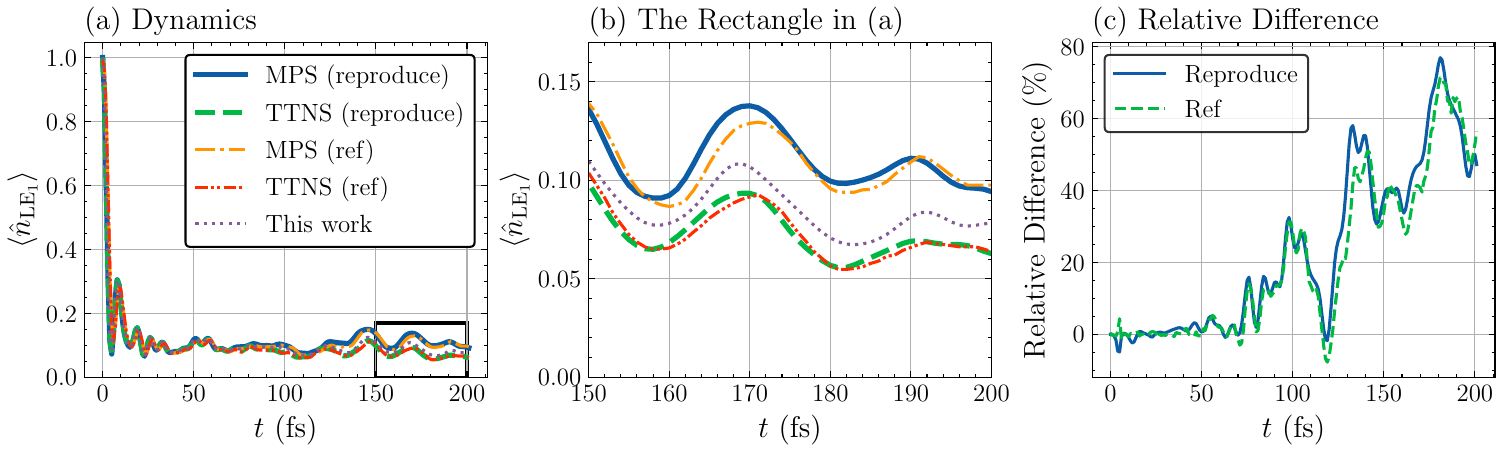}
    \caption{Reproduction of the \rev{difference between MPS and TTNS} in the reference paper~\cite{dorfner2024comparison}. (a) The dynamics of $\braket{\hat n_{\textrm{LE}_1}}$ calculated using Chain and Tree. \rev{(b) The dynamics at the long time limit, i.e., from $150$ to $200$ fs.} (c) The relative difference between Chain and Tree. ``Reproduce'' represents the results calculated by us. ``Ref'' represents the data extracted from the reference paper. \rev{``This work'' represents the converged dynamics reported in the main text.}}
    \label{fig:reproduce}
\end{figure}

Using our \textrm{Renormalizer} implementation, we successfully reproduce 
the \rev{difference between Chain and Tree} structures
reported in the reference work~\cite{dorfner2024comparison}.
Fig.~\ref{fig:reproduce} demonstrates that our reproduced results and the original data points extracted from the reference work are in excellent agreement. 
This validation confirms both the reliability of our implementation and the reproducibility of the earlier findings.
\rev{Besides, from Fig.~\ref{fig:reproduce}(b), the Tree simulation in the reference paper agrees better with the results in this work compared with the Chain simulation.}

\begin{figure}[htbp]
    \centering    \includegraphics[width=.9\textwidth]{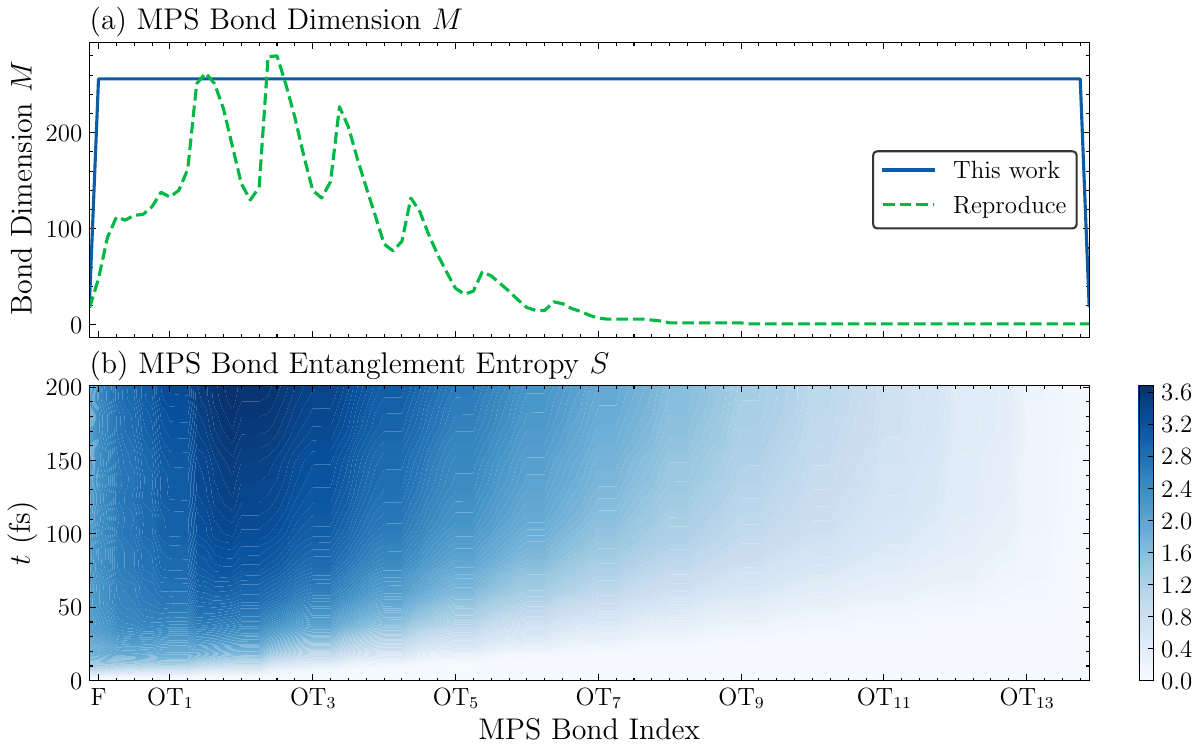}
    \caption{Bond dimension and entanglement entropy analysis for Chain calculations. (a) Comparison of bond dimension distributions between this work and the implementation by expanding bond dimension using 2-site TDVP-PS. (b) The numerically exact bond entanglement entropy profile. In (a), ``this work'' means the bond dimension distribution employed in the calculations in the main text. And ``Reproduce'' means the bond dimension distribution employed in Fig.~\ref{fig:reproduce}.}
    \label{fig:entanglement-m}
\end{figure}

The key factor enabling this reproduction is the bond dimension setup. While maintaining nearly identical computational parameters to those used in our main text, where we achieved 10\% relative difference \revrev{and 0.005 absolute difference}, we specifically matched the bond dimension of the reference work. 
More specifically, for MPS calculation, starting from a Hartree product state where $M=1$, we perform 2-site TDVP-PS for a few steps, 
with a very small singular value truncation threshold of $10^{-6.5}$.
While this procedure allows rapid growth of the bond dimension,
it creates an uneven distribution where regions of initially strong entanglement develop large bond dimensions while other areas remain constrained.
When the maximum bond dimension is above a target value, which in this case is 250, the time evolution algorithm is switched to 1-site TDVP-PS, and the time evolution proceeds with fixed bond dimension.
The transition happens at very early stages of the time evolution ($t<20$ fs).
A comparison of the bond dimension obtained using this approach and the fixed uniform bond dimension employed in this work is included in Fig.~\ref{fig:entanglement-m}(a).
The direct result of this setup is that the region where the initial entanglement is small has constrained bond dimension \rev{throughout} the time evolution.
In Fig.~\ref{fig:entanglement-m}(b), we plot the accurate bipartite entanglement entropy at each bond using our uniform bond dimension.
We can see  that the region where initially has small entanglement developes moderate amount of entanglement entropy
at later times.
The small bond dimension in this region creates persistent bottlenecks in the calculation and limits the overall accuracy regardless of maximum bond dimension elsewhere.
Increasing the maximum bond dimension has very slow convergence because the maximum bond dimension is increasing much faster over time than the bond dimension in the low-entanglement region.

On the other hand, for the Tree structure calculation, we replicate the tree structure and bond dimension reported in the reference work, including its maximum bond dimension of 40.
As shown in Fig.~\ref{fig:ext}, $M=40$ \rev{will underestimate the occupation in the long times},
which is consistent with the underestimation in Fig.~\ref{fig:reproduce}(a).
The combined inaccuracies from both MPS and TTNS implementations account for the observed discrepancy in the reference work.

\begin{figure}[htbp]
    \centering    \includegraphics[width=\textwidth]{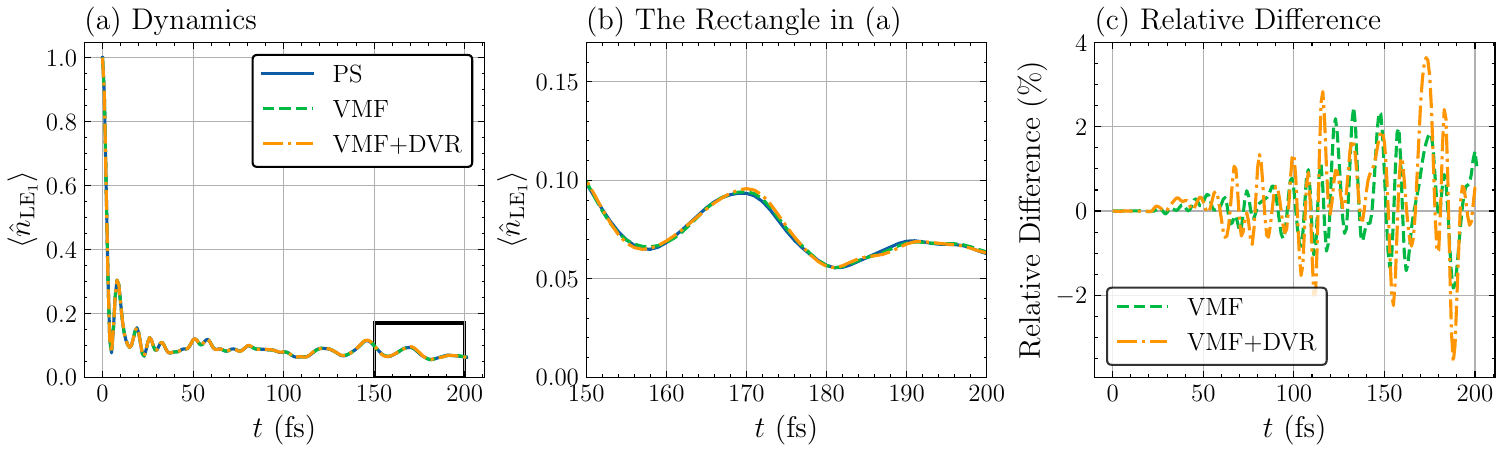}
    \caption{The effect of time evolution scheme and primitive basis on the time evolution of $\braket{\textrm{LE}_1}$. (a) The full dynamics from 0 to 200 fs. (b) The dynamics from $150$ to $200$ fs.
    (c) The relative difference between the $\braket{\hat n_{\textrm{LE}_1}}$ values, using PS values as the reference.}
    \label{fig:vmf-dvr}
\end{figure}

\begin{figure}[h]
    \centering    \includegraphics[width=\textwidth]{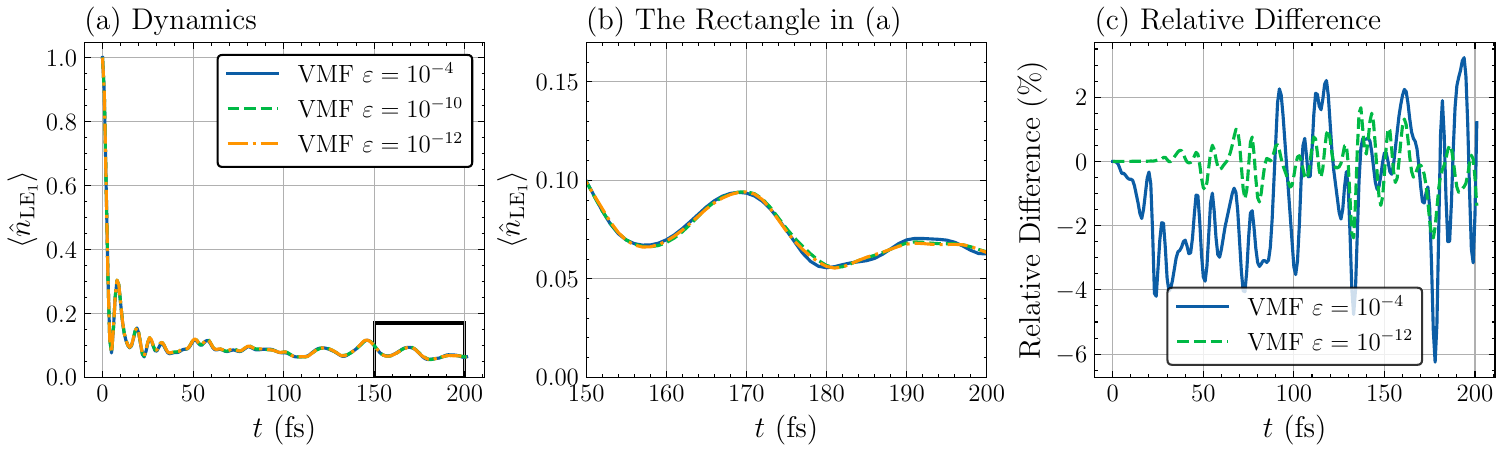}
    \caption{\rev{The effect of different regularization parameter $\varepsilon$ to the dynamics. (a) The dynamics of $\braket{\hat n_{\textrm{LE}_1}}$ calculated using different $\varepsilon$. \rev{(b) The dynamics at the long time limit, i.e., from $150$ to $200$ fs.} (c) The relative difference between $\varepsilon=10^{-4}$ and $\varepsilon=10^{-10}$, and between $\varepsilon=10^{-12}$ and $\varepsilon=10^{-10}$.}}
    \label{fig:regularization}
\end{figure}

To investigate the potential effect of methodological differences beyond bond dimension, we examine two additional factors from the reference study.
The first is the use of different time evolution schemes.
\revrev{The reference work employs the PS method for MPS and the VMF~\cite{beck2000multiconfiguration} scheme for TTNS, as VMF is the traditional integrator for ML-MCTDH}.
To investigate their effect, we perform further benchmarks based on the Tree structure and bond dimension in Fig.~\ref{fig:vmf-dvr}.
We find that the VMF and PS schemes show excellent agreement, with $\sim2\%$ \revrev{relative} difference \revrev{(0.001 absolute difference)},
which is consistent with our previous MPS benchmarks ~\cite{li2020numerical}.
Besides, in the reference work, the harmonic oscillator eigenbasis and DVR basis are employed in MPS and TTNS calculation respectively.
We perform VMF time evolution with sine DVR basis~\cite{beck2000multiconfiguration}
and the results are also included in Fig.~\ref{fig:vmf-dvr}.
We find that the two factors collectively \rev{contribute} to 3\% of relative difference \revrev{(0.002 of absolute difference)}.
In Fig.~\ref{fig:regularization} we show the effect of different regularization parameter $\varepsilon$.
In our VMF integration the density matrix $\rho$ is regularized using the following equation
\begin{equation}
    \rho' = \rho + \varepsilon \exp{-\rho/\varepsilon}
\end{equation}
where by default $\varepsilon=10^{-10}$.
As shown in in Fig.~\ref{fig:regularization}, using a small $\varepsilon$ will lead to larger error in the initial dynamics.
In the long time limit, the difference caused by $\varepsilon$ is smaller than 6\% \revrev{(0.003 of absolute difference)}.
We conclude that neither factor significantly contributes to the discrepancy observed in the reference work, and that bond dimension remains the primary determinant of accuracy in these calculations.

\section{Extrapolated Dynamics}
\label{sec:all-dof}
\begin{figure}[htbp]
    \centering    \includegraphics[width=.95\textwidth]{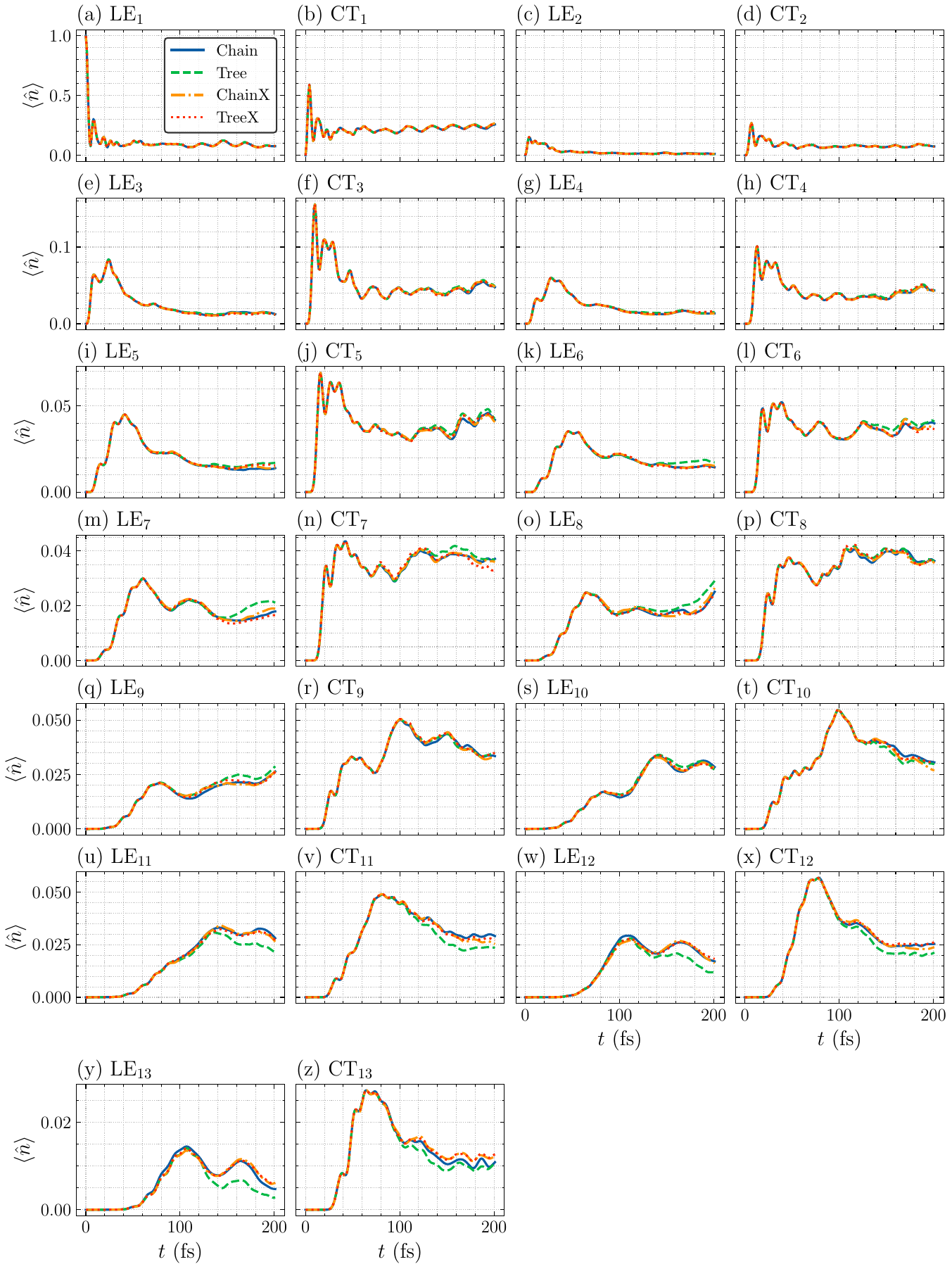}
    \caption{The extrapolated occupation dynamics for all electronic states by \rev{Chain, Tree, ChainX and TreeX.}}
    \label{fig:full-dynamics}
\end{figure}

In this section, we show the extrapolated dynamics of all 26 electronic states in Fig.~\ref{fig:full-dynamics}.
The results by \rev{Chain, Tree, ChainX and TreeX} are all shown in each panel.
The excellent agreement across all states validates the consistency of \rev{MPS and TTNS}, with most cases showing nearly indistinguishable curves between methods.
While minor deviations appear for certain states, such as in Fig.~\ref{fig:full-dynamics}(m),
these occur primarily where the occupation is small in magnitude.
For example, in Fig.~\ref{fig:full-dynamics}(m), LE$_7$ shows visible difference between Tree and TreeX, yet the absolute difference is \rev{approximately 0.004}.

\section*{Acknowledgement}
Weitang Li and Jun Yan are supported by the Guangdong Basic Research Center of Excellence for Aggregate Science
and
the Shenzhen Science and Technology Program (No. KQTD20240729102028011). 
Jiajun Ren is supported by the National Natural
Science Foundation of China (Grant No. 22273005 and No. 22422301).

\section*{Code Availability}
The complete code that supports reproducing all data in this study is deposited on Zenodo~\cite{data}.

\section*{Data Availability}
The source data for all figures and the Python scripts to produce all figures are deposited in Zenodo~\cite{data}.

\section*{Competing interests}
The authors declare no competing interests.


\bibliography{refs}

\clearpage

\begin{figure}[htp]
    \centering
    \includegraphics[width=0.5\textwidth]{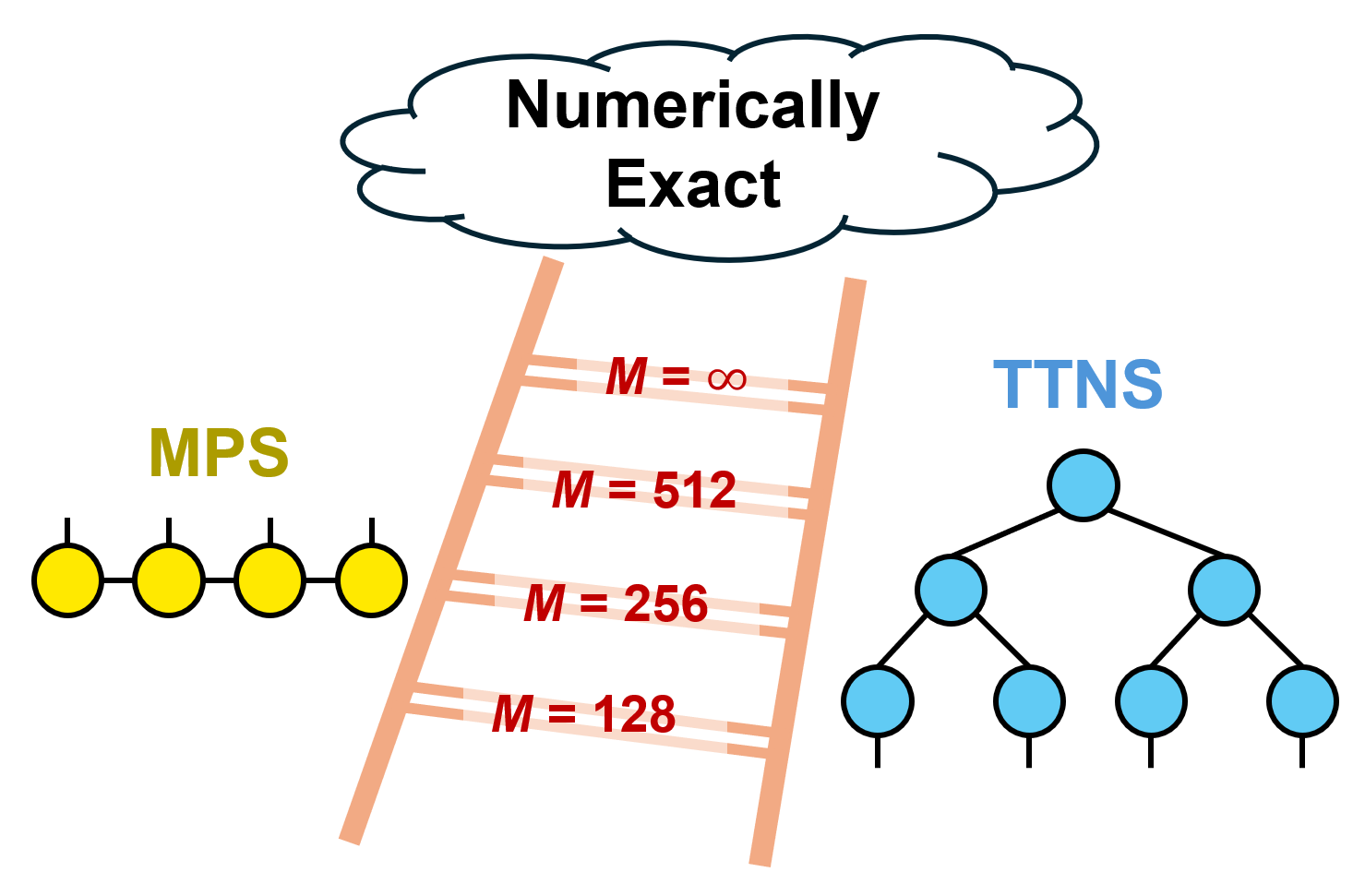}
\caption*{For Table of Contents Only}
\end{figure}

\end{document}